\documentclass[a4paper,12pt,preprint,amsmath,showpacs,
nofootinbib,superscriptaddress]{revtex4}
\usepackage{mathrsfs}
\usepackage{amssymb}
\usepackage{amsmath}
\usepackage{graphicx}
\usepackage{multirow}
\usepackage{subfigure}
\usepackage{float}
\usepackage{graphics}

\def\as{\alpha_s}

\def\nno{\nonumber\\}
\newcommand{\be}{\begin{eqnarray}}
\newcommand{\ee}{\end{eqnarray}}

\def\ses{\sqrt{S}=7~\text{TeV}}
\def\eis{\sqrt{S}=8~\text{TeV}}

\def\fos{\sqrt{S}=14~\text{TeV}}
\def\mvv{M_{V_1V_2}}
\def\zz{ZZ}
\def\ww{W^{+}W^{-}}
\def\wz{W^{\pm}Z}
\def\ptv{p_T^{veto}}
\def\pis{\pi^2~\text{enhancement effects}}
\def\gev{\text{GeV}}
\def\tev{\text{TeV}}
\def\nlonnll{\text{NLO + NNLL}}
\def\muhs{\mu_h^2}

\begin{document}
\title{Resummation prediction on gauge boson pair production with a jet veto}

\author{Yan  Wang}
\email{yancywww@pku.edu.cn}
\affiliation{Department of Physics and State Key Laboratory of
Nuclear Physics and Technology, Peking University, Beijing 100871,
China}
\author{Chong Sheng Li}
\email{csli@pku.edu.cn}
\affiliation{Department of Physics and State Key Laboratory of
Nuclear Physics and Technology, Peking University, Beijing 100871,
China}
\affiliation{Center for High Energy Physics, Peking University, Beijing 100871, China}
\author{Ze Long Liu}
\email{liuzelong@pku.edu.cn}
\affiliation{Department of Physics and State Key Laboratory of
Nuclear Physics and Technology, Peking University, Beijing 100871,
China}




\begin{abstract}
We investigate the resummation effects with a jet veto, for $\wz$ and $\zz$ productions at the LHC in soft-collinear effective theory.
We present the invariant mass distributions and the total cross section with different jet veto $\ptv$ and jet radius $R$ for these process at Next-to-Next-to-Leading- Logarithmic level.
Our results show that the jet-veto resummation can increase the jet-veto cross section and decrease the scale uncertainties, especially in the large center-of-mass energy.
We find that for $\ptv>30~\gev$ and $R=0.4$, the resummation results can increase POWHEG+PYTHIA predictions by about $19\%$ for $\wz$ production and $18\%$ for $\zz$ production, respectively.
Our results agree with the CMS data for $\wz$ productions within $2\sigma$ C.L. at $\eis$, which can explain the 2$\sigma$ discrepancy between the  CMS experimental results and theoretical predictions based on NLO calculation with parton showers.
\end{abstract}

\maketitle
\section{INTRODUCTION}\label{s1}
It is particularly important  to precisely test the Standard Model (SM) electroweak sector in the $\ww$, $\wz$ and $\zz$ productions, which are also significant backgrounds for measuring Higgs boson property and searching for new physics. So it is  vital to understand the productions of gauge boson pair well.

The high order QCD corrections to the gauge boson pair productions have been studied for a long time~\cite{PhysRevD.43.3626,Mele:1990bq,PhysRevD.50.1931,PhysRevD.60.114037,PhysRevD.60.113006,Dixon19983,Frixione:1992pj,Campanario:2012fk,Cascioli:2014yka,Gehrmann:2013cxs,Henn:2014lfa,Gehrmann:2014bfa,Caola:2014lpa}.
Some resummation effects are also calculated in recent years~\cite{Wang:2013qua,Dawson:2013lya,Wang:2014mqt}.
ATLAS and CMS collaborations at the LHC have made much effort on measuring $\ww$, $\wz$ and $\zz$ productions ~\cite{Aad:2011xj,Aad:2012twa,Aad:2011cx,ATLAS-CONF-2014-015,ATLAS-CONF-2013-021,ATLAS-CONF-2013-020,Chatrchyan:2012sga,CMS-PAS-SMP-12-016,CMS-PAS-SMP-13-005,CMS-PAS-SMP-13-011,Chatrchyan:2014aqa}.
The ATLAS and CMS measurements for $\zz$ productions are in remarkable agreement with the SM NLO predictions. However, the experimental results are deviated from the NLO predictions for $\ww$ production over 2$\sigma$ in both ATLAS and CMS measurements at $8~\tev$, and there is  also about 2$\sigma$ discrepancy for $\wz$ pair productions in the CMS measurements.

In these experiments, a jet veto is often imposed to increase the ability to suppress backgrounds. When jet-veto efficiencies ($\epsilon(\ptv)\equiv \sigma(\ptv)/\sigma_{inclusive}$) are simulated, the inclusive cross section can be obtained, where $\sigma(\ptv)$ is the jet-veto cross sections.
Since a small energy scale of the jet veto $\ptv$ is introduced, which is at the order of 20-30 GeV, there will be a large logarithmic terms  $\ln\ptv/Q$ in the perturbative calculations at  all orders, where Q denotes the hard scale in the process.
These large logarithms need to be resummed for improving the accuracy of the theoretical predictions.
In the experimental analysis, POWHEG+PYTHIA~\cite{Alioli:2010xd} are always used to provided approximate NLL resummation for obtaining $\epsilon(\ptv)$. But the parameters in these softwares need to be tuned finely, which are considered as an explanation for the deviation of the measured data with the NLO predictions in $\ww$ channel, where the matched parton shower are found to overestimate the Sudakov suppression and lead to lower theoretical predictions~\cite{Monni:2014zra}.
In~\cite{Jaiswal:2014yba}, it was claimed that the discrepancy in $\ww$  channel can be explained by the resummation with a jet veto at the $\nlonnll$ accuracy. In the newest CMS report, the $\ww$ cross section is consistent with the NNLO theoretical prediction~\cite{CMS-PAS-SMP-14-016}.  Therefore, higher order jet-veto resummations for $\wz$ and $\zz$ productions need to be considered, since NNLO prediction for $\wz$ production is still hard to obtain .

In Ref.~\cite{Becher:2014aya}, an automated method to perform resummations for vector-boson productions involving jet vetoes are proposed in the framework of SCET. Results of $\ww$ production is calculated as an example, which are in agreement with the measurements at the LHC, but the results for $\wz$ and $\zz$ channel are not available~\cite{Becher:2014aya}.

In this paper, we present the calculations of the resummation for $\wz$ and $\zz$ production processes with jet vetoes at the $\nlonnll$ level in SCET, which are compared with the POWHEG+PYTHIA simulation. Finally, we try to explain the discrepancy of $\wz$ channel with the CMS measuremental data.

The paper is organized as follows. In Sec.~\ref{s2}, we describe the formalism for jet-veto resummation in SCET briefly. In Sec.~\ref{s3}, we present the numerical results and some discussion. Then Sec.~\ref{s4} is a brief conclusion.

\section{RG improved cross section}\label{s2}
The gauge boson pair production cross section defined with a jet veto $\ptv$ can be resummed in SCET. The interested kinematic region is
\be
\hat{s},\mvv^2,m_V^2 \gg {\ptv}^2 \gg \Lambda_{QCD}^2.
\ee
where $V$ stands for $W^{\pm}$ or $Z$ boson. We can obtained the factorization formula by factorizing the contributions of hard, collinear, anti-collinear, and soft degrees of freedom in SCET.
\be
\frac{d\sigma }{dy}=\sigma _0\mathcal{H}_{VV}\left(-q^2,\mu \right)
   \mathfrak{B}_c\left(\xi _1,\ptv,\mu \right) \mathfrak{B}_{c'}\left(\xi _2,\ptv,\mu \right)\mathcal{S}\left(\ptv,\mu \right)
\ee
where $\xi_{1,2}=(\mvv/\sqrt{s})e^{\pm y}$, and $\mathcal{H}_{VV}$ is the hard function, which can be expand in terms of $\alpha_s$.
\be
 \mathcal{H}_{VV}= \mathcal{H}_{VV}^{(0)} + \frac{\alpha_s}{4\pi}\mathcal{H}_{VV}^{(1)} + \cdots.
\ee
The expression of $\mathcal{H}_{VV}$ can be found in Ref.~\cite{Wang:2013qua}. The hard function satisfies the renormalization-group (RG) equation
\be\label{s2_eq_rg_h}
\frac{d}{d\ln\mu} \mathcal{H}_{VV}\left(M,\mu \right)= 2\left[ \Gamma_{\rm cusp}^F(\alpha_s)\ln\frac{-M^2}{\mu^2} + 2\gamma^{q}(\alpha_s)\right]\mathcal{H}_{VV}\left(M,\mu \right).
\ee
As in the case of transverse momentum, there needs collinear anomalous term, $F_{cc'}$, to cancel large logarithms of the scale ratio $\mvv/\ptv$
\be
\mathfrak{B}_c\left(\xi _1,\ptv,\mu \right) \mathfrak{B}_{c'}\left(\xi _2,\ptv,\mu \right)\mathcal{S}\left(\ptv,\mu \right) = \quad\quad\quad\quad\quad\quad\quad\quad\nno
\quad\quad\quad \left(\frac{M^2}{{\ptv}^2}\right){}^{-F_{\text{cc}'}\left(\ptv,\mu \right)} e^{2 h_F\left(\ptv,\mu \right)} \bar{B}\left(z_1,\ptv,\mu \right) \bar{B}\left(z_2,\ptv,\mu \right).
\ee
$\bar{B}_c$ is RG invariant beam function, which can be matched to the standard PDFs as\cite{Becher:2012qa}
\be
\bar{B}_c\left(\xi ,x_T^2,\mu \right)=\sum_i\int \frac{dz}{z}\phi _c\left(\xi/z,\mu \right) \bar{I}_{i\leftarrow j}\left(z,x_T^2,\mu \right)
\ee
The function $\bar{I}_{\text{qi}}(z,\ptv,\mu)$ has already be computed at one-loop order\cite{Becher:2013xia}:
\be
\bar{I}_{qi}(z,\ptv,\mu)&=&\delta _{qi} \delta (1-z)-\frac{1}{2}\alpha_s L  P_{qi}(z)+\alpha _s  R_{qi}(z)
\ee

The RG variant term of the beam function are factorized out as the exponents $h_F$, its RG equations is\cite{Becher2012}
\be
\frac{d}{d \ln\mu} h_F(\ptv,\mu) = 2\Gamma_{cusp}^F \ln\frac{\mu}{\ptv} - 2\gamma^q(\mu)
\ee
Solving the equation, we obtain
\be
h_1&=&\alpha _S \left(\frac{1}{4} L^2 \Gamma _0^F-L \gamma
   _0^q\right)\nno
h_2&=&\alpha _S^2 \left(\frac{1}{12} \beta _0 L^3 \Gamma_0^F+\frac{1}{4} L^2 \left(\Gamma _1^F-2 \beta _0 \gamma_0^q\right)-L \gamma _1^q\right)\nno
h_F&=&h_1+h_2
\ee
$F_{cc'}$ obey the RG equations
\be
\frac{d}{d ln\mu} F_{cc'}(\ptv,\mu) = 2\Gamma_{cusp}^F(\mu)
\ee
Solving and writting it as the terms of $\alpha_s$, we obtain
\be
F_{cc'}=\alpha_s L  \Gamma _0^F  + \alpha _S^2 \left(d_2^{\text{veto}}(R)+\frac{1}{2} \beta_0 L^2 \Gamma _0^F + L \Gamma _1^F\right)
\ee
The two loop coefficient $d_2^{veto}(R)$ has the form\cite{Becher:2013xia}
\be
d_2^{veto} =  d_2^q - 8 \Gamma_0^F f(R),
\ee
where $d_2^q$ is obtained from the small transverse momentum resummation for Drell-Yan process\cite{Becher:2013xia}:
\be
d^q_2 = \Gamma^F_0\left[\left(\frac{202}{27} -7\xi\right) C_A -\frac{56}{27}T_F n_f\right]
\ee
The function $f(R)$ can be found in Ref~\cite{Becher:2012qa} as
\be
f(R) &=& -(1.09626 C_A +0.1768 n_f T_F ) \ln R +(0.6072C_A - 0.0308T_F n_f) \nno
&& +(0.2639C_A - 0.8225C_F + 0.02207T_F n_f)R^2\nno
 && - (0.0226C_A - 0.0625C_F + 0.0004T_F n_f)R^4 + \ldots.
\ee

The RG evolving terms can be written into an exponent factor $g_F$
\be
g_F=-F_{cc'} \left(L+\ln \left(\frac{M^2}{\mu
   ^2}\right)\right)+2 h_F.
\ee

So
\be
 && \mathfrak{B}_c\left(\xi _1,\ptv,\mu \right) \mathfrak{B}_{c'}\left(\xi _2,\ptv,\mu \right)\mathcal{S}\left(\ptv,\mu \right)  \nno
&& = \sum_{i,j} \int_{z_1}^1 \int_{z_2}^1 \frac{d z_1}{z_1} \frac{d z_2}{z_2}  e^{g_F} \bar{I}\left(z_1,\ptv,\mu \right) \bar{I}\left(z_2,\ptv,\mu \right) f_i(\xi_1/z_1,\mu)f_j(\xi_2/z_2,\mu).
\ee
The total cross section can be rewritten as
\be
\frac{d\sigma }{dy} &=& \sigma _0 \mathcal{H}_{VV}\left(-M_{VV}^2,\mu \right)
   \sum_{i,j} \int_{z_1}^1 \int_{z_2}^1 \frac{d z_1}{z_1} \frac{d z_2}{z_2}\nno
    && e^{g_F} \bar{I}\left(z_1,\ptv,\mu \right) \bar{I}\left(z_2,\ptv,\mu \right) f_i(\xi_1/z_1,\mu)f_j(\xi_2/z_2,\mu).
\ee
In addition to the singular terms, we should also contain contributions from the non-singular terms, which can be obtained by matching resummed results to the full fixed order cross section. Finally ,the RG improved prediction for the gauge boson pair can be expressed as
\be\label{s2_eq_fullcs}
\frac{d\sigma^{\rm \nlonnll}}{d\mvv^2}=\frac{d\sigma^{\rm NNLL}}{d\mvv^2} +\left(\frac{d\sigma^{\rm NLO}}{d\mvv^2}-\frac{d\sigma^{\rm NNLL}}{d\mvv^2}\right)_{\rm expanded~to~NLO}.
\ee
\section{Numerical results}\label{s3}
In this section, we present the numerical results for the jet-veto resummation effects on gauge boson pair productions. We choose SM input parameters as~\cite{Beringer:1900zz}:
\begin{eqnarray}\label{sm_para}
 &&  m_W= 80.4 \textrm{~GeV}, \quad    m_Z = 91.19 \textrm{~GeV}, \quad \alpha(m_Z)=1/132.338.
\end{eqnarray}
MSTW2008nnlo PDF sets and the corresponding running QCD coupling constant are used throughout the paper.
Unless specified otherwise, we set $\mu_h$ at the scales of the hard scattering process $\mu_h^2\sim\mvv^2$ to minimize $\log(-\mvv^2/\mu_h^2)$ logarithms.
And we choose $\mu_h^2=-\mvv^2$ to reduce the $\pi^2$ terms arising from the square of the $\log(-1)$ to improves the perturbative convergence.
The NLO QCD corrections are calculated by  Monte Carlo for FeMtobarn processes (MCFM)~\cite{PhysRevD.60.113006}, where  factorization and renormalization scales are chosen as $\mvv$.
\subsection{Leading singular terms}
\begin{figure}[t]
\begin{minipage}{0.49\linewidth}
\centering
  \includegraphics[width=1.0\linewidth]{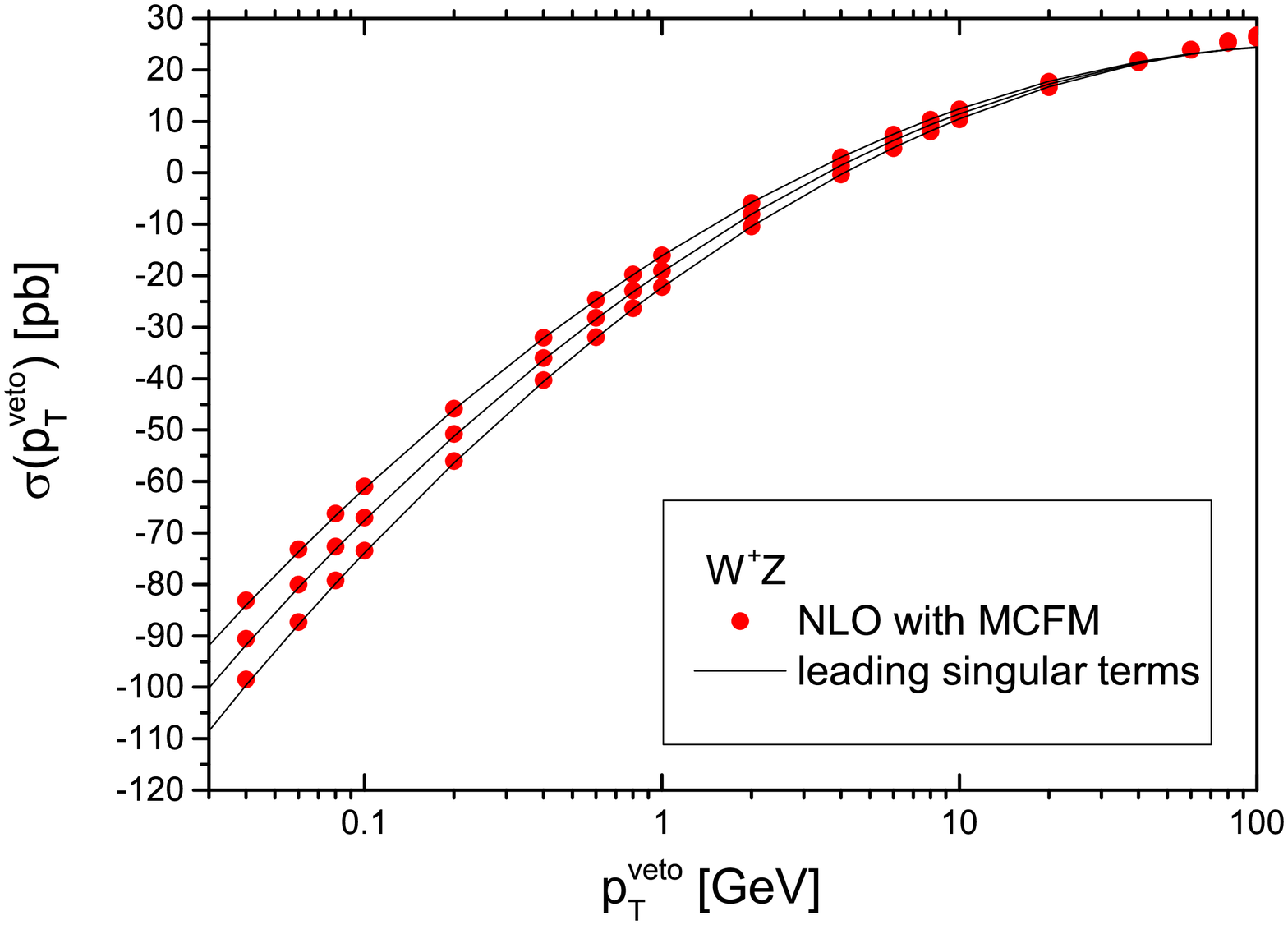}\\
\end{minipage}
\hfill
\begin{minipage}{0.49\linewidth}
\centering
 \includegraphics[width=1.0\linewidth]{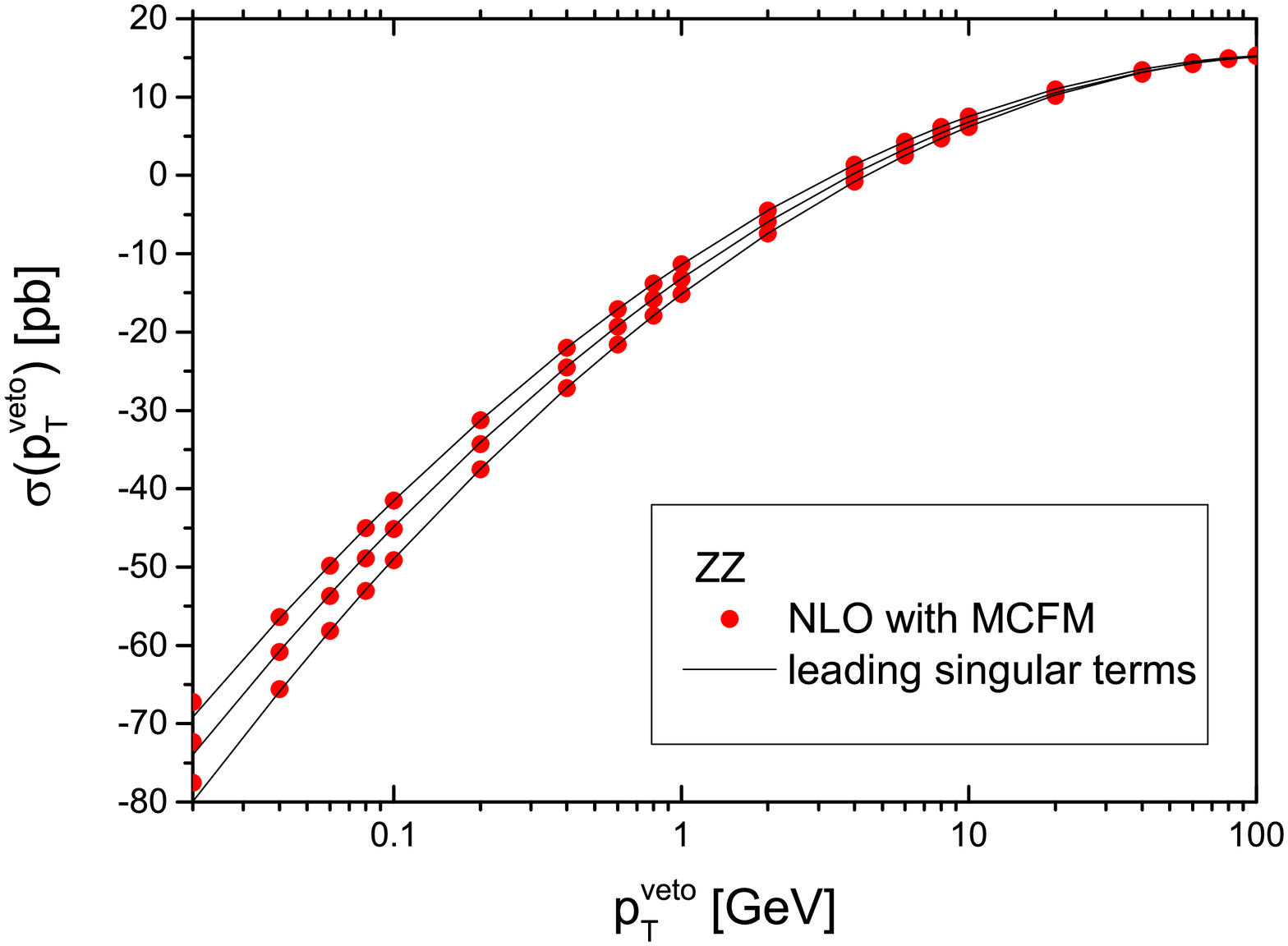}\\
\end{minipage}
\caption{Comparisons of the leading singular and the exact NLO jet vetoed cross sections for $W^{+}Z$ (left panel) and $\zz$ (right panel) production at the LHC with $\fos$, respectively.}\label{f_match}
\end{figure}
We expand the resummed results to the leading singular term to verify its correctness. In Fig~\ref{f_match}, we compare them with the exact NLO results calculated by  MCFM. We can see that the leading singular terms of the cross section with jet vetoes can reproduce the exact NLO jet vetoed cross section in the small $\ptv$ region for both $\wz$ and $\zz$ productions.

\subsection{Scale dependence}
\begin{figure}[t]
\begin{minipage}{0.49\linewidth}
\centering
  \includegraphics[width=1.0\linewidth]{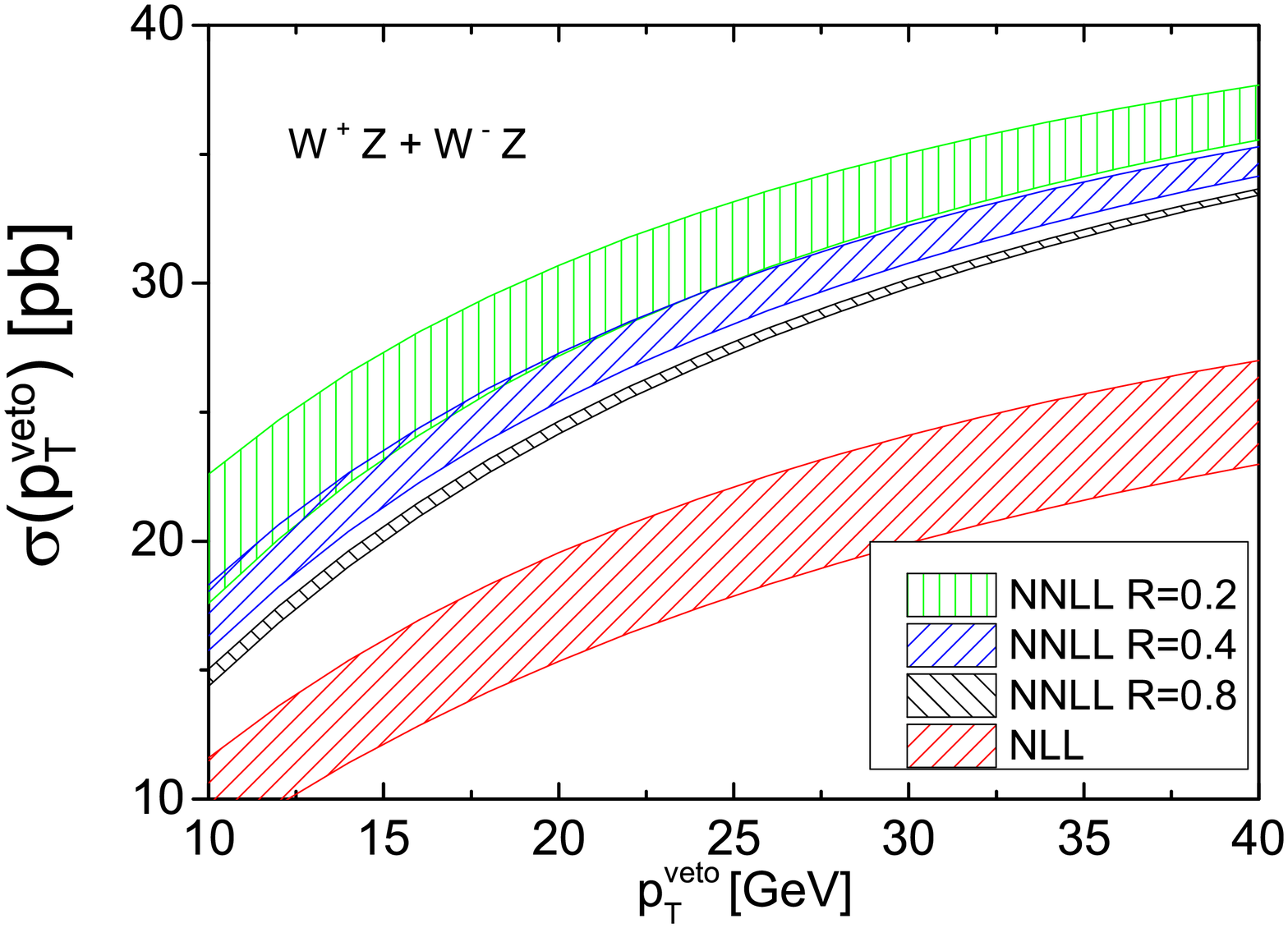}\\
\end{minipage}
\hfill
\begin{minipage}{0.49\linewidth}
\centering
 \includegraphics[width=1.0\linewidth]{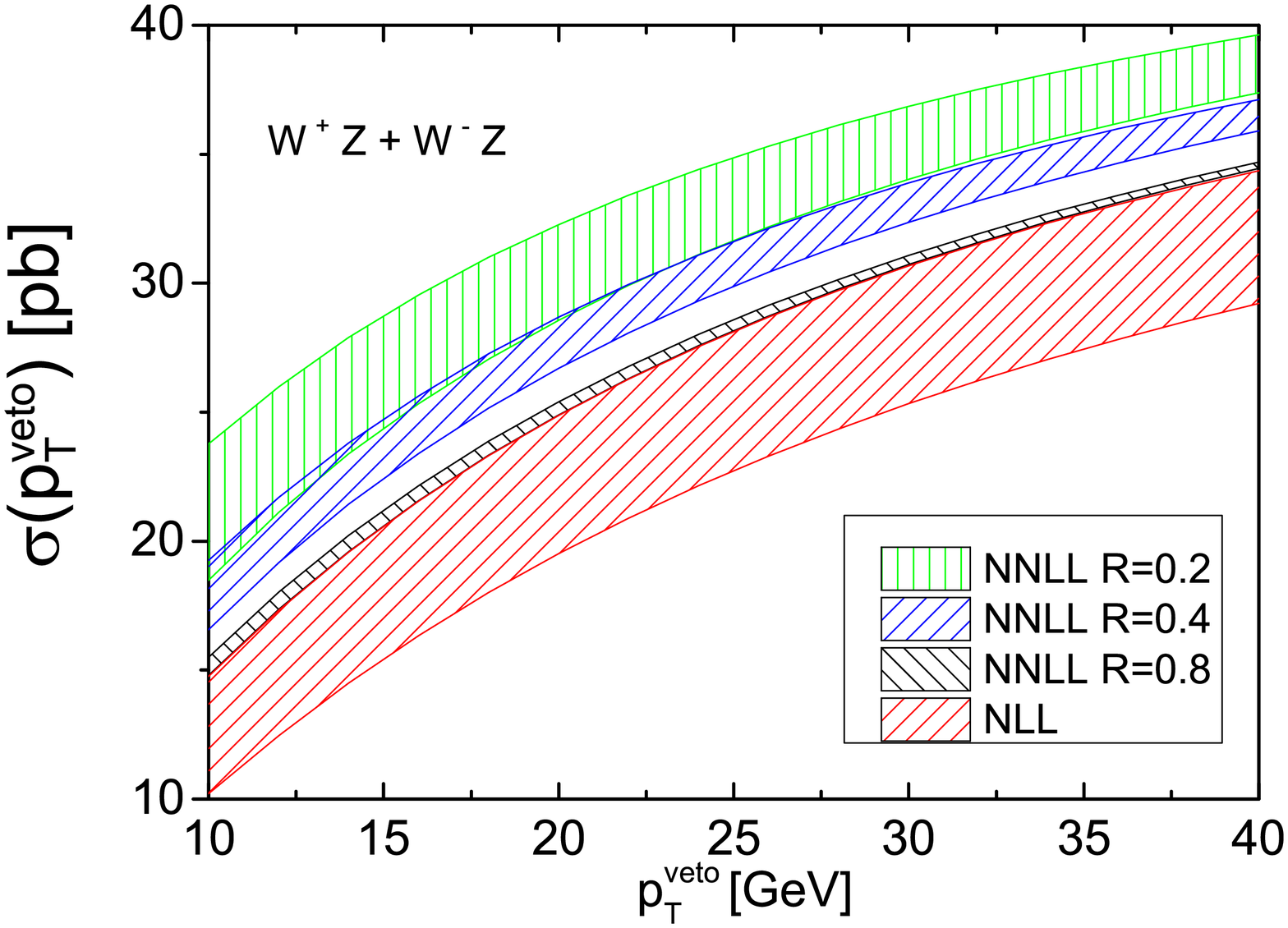}\\
\end{minipage}
\caption{The NLL and NNLL  resummed jet veto cross section for $\wz$ production  with $\fos$ at the LHC. The right figure is the same as the left one except that it is with $\pis$.}\label{f_NNLL_scale_wz_s2}
\end{figure}

\begin{figure}[t]
\begin{minipage}{0.49\linewidth}
\centering
  \includegraphics[width=1.0\linewidth]{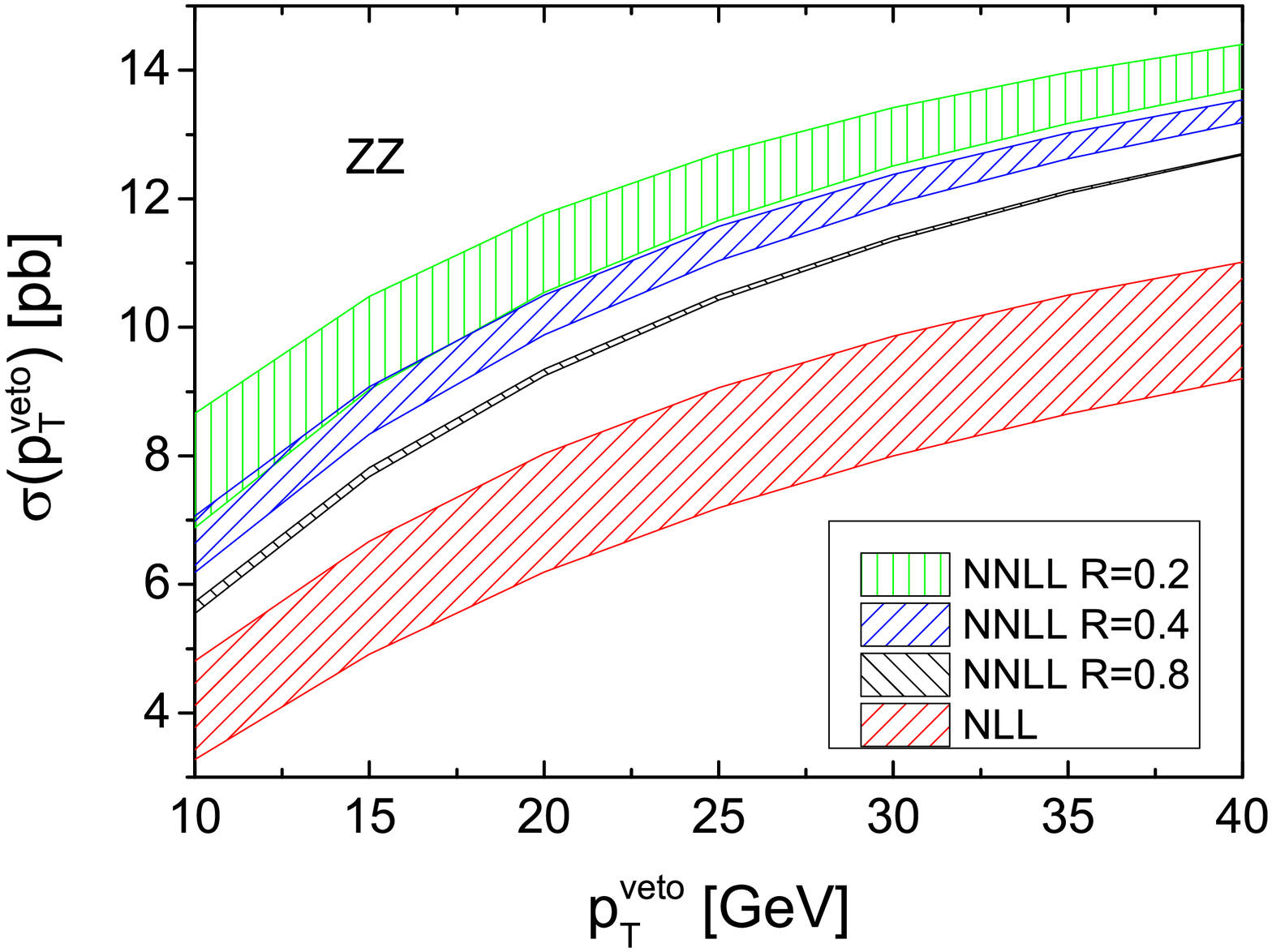}\\
\end{minipage}
\hfill
\begin{minipage}{0.49\linewidth}
\centering
 \includegraphics[width=1.0\linewidth]{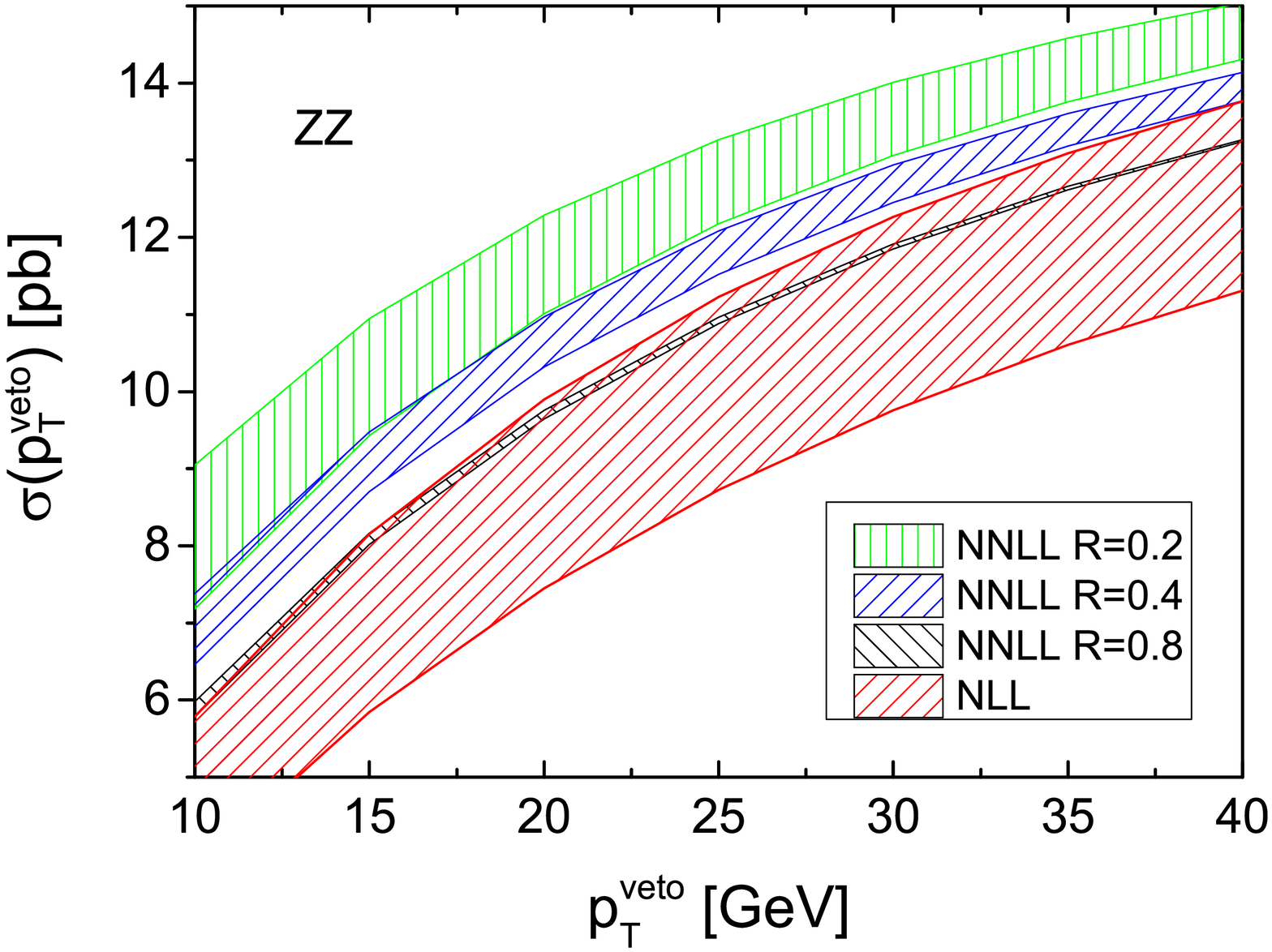}\\
\end{minipage}
\caption{The NLL and NNLL  resummed jet veto cross section for $\zz$ production  with $\fos$ at the LHC. The right figure is the same as the left one except that it is with $\pis$.}\label{f_NNLL_scale_zz_s2}
\end{figure}

In Fig.~\ref{f_NNLL_scale_wz_s2} and Fig.~\ref{f_NNLL_scale_zz_s2},
we show the scale dependence on $p_T^{veto}$ at NLL and NNLL for different jet radius R when $\fos$, where the hard scale are fixed as $\mu_h^2=\mvv^2$ and the error bands only reflect the scale uncertainties by varying the scales in the range $\ptv/2<\mu_f<2~\ptv$.
Unless specified otherwise, we will follow the same choice as above, because the uncertainties from the hard-scale variation are very small.
From Fig.~\ref{f_NNLL_scale_wz_s2} and Fig.~\ref{f_NNLL_scale_zz_s2}, we can see that the scale dependence of the NNLL results are much smaller than the NLL one.
In resummed results, R dependent term has the form $\exp(\as^2 d_2^{veto}(R) \ln(M/{\ptv}))$. Therefore   scale dependence decreases rapidly along with the increase of R; and for a fixed R, when $\ptv$ increases, it becomes smaller slowly, which are all reflected in those figures.
The results including $\pis$  are also presented in the Fig.~\ref{f_NNLL_scale_wz_s2} (right) and Fig.~\ref{f_NNLL_scale_zz_s2} (right), where the theoretical convergence are improved obviously.

\begin{figure}[t]
\begin{minipage}{0.49\linewidth}
\centering
  \includegraphics[width=1.0\linewidth]{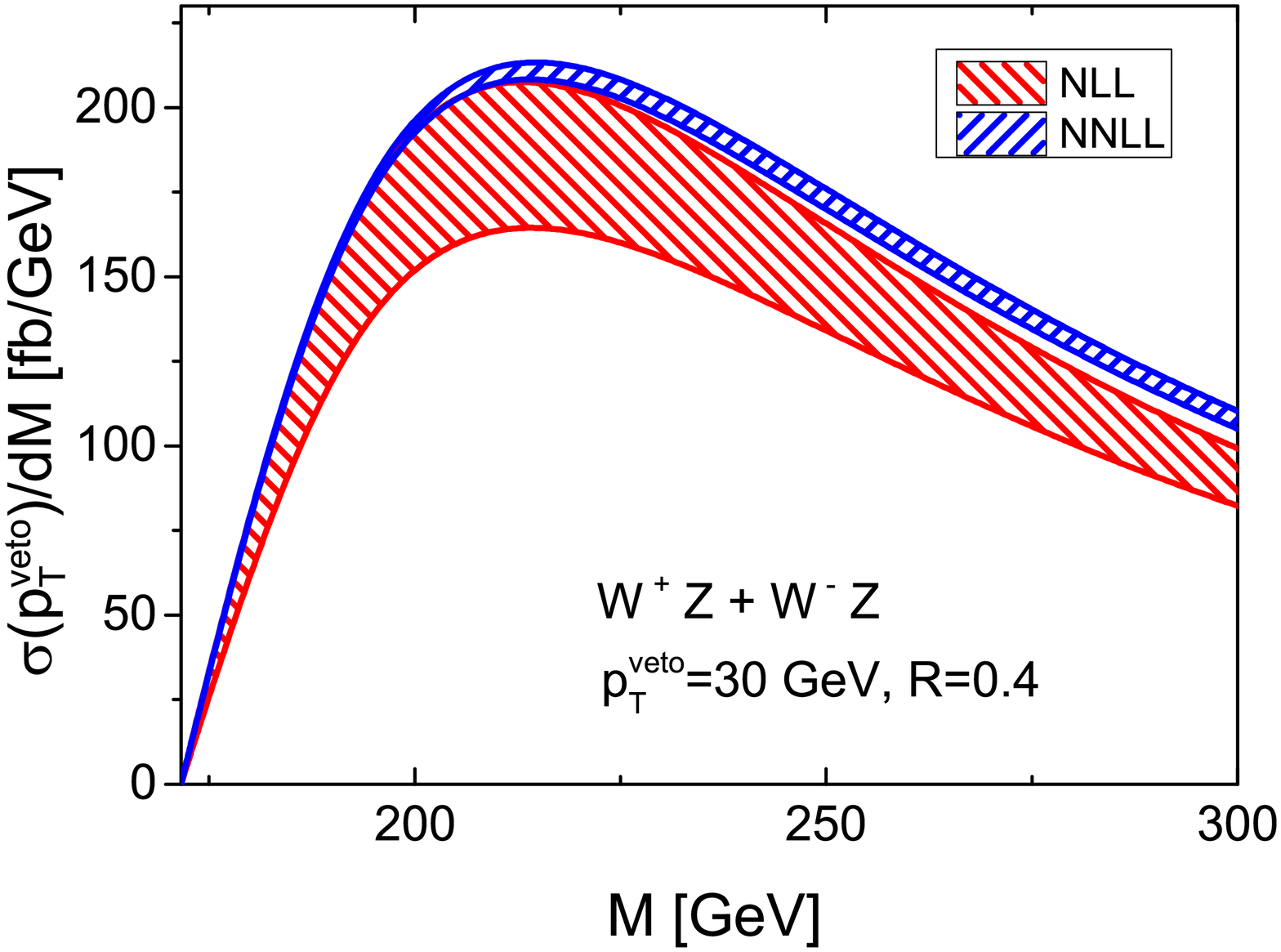}\\
\end{minipage}
\hfill
\begin{minipage}{0.49\linewidth}
\centering
 \includegraphics[width=1.0\linewidth]{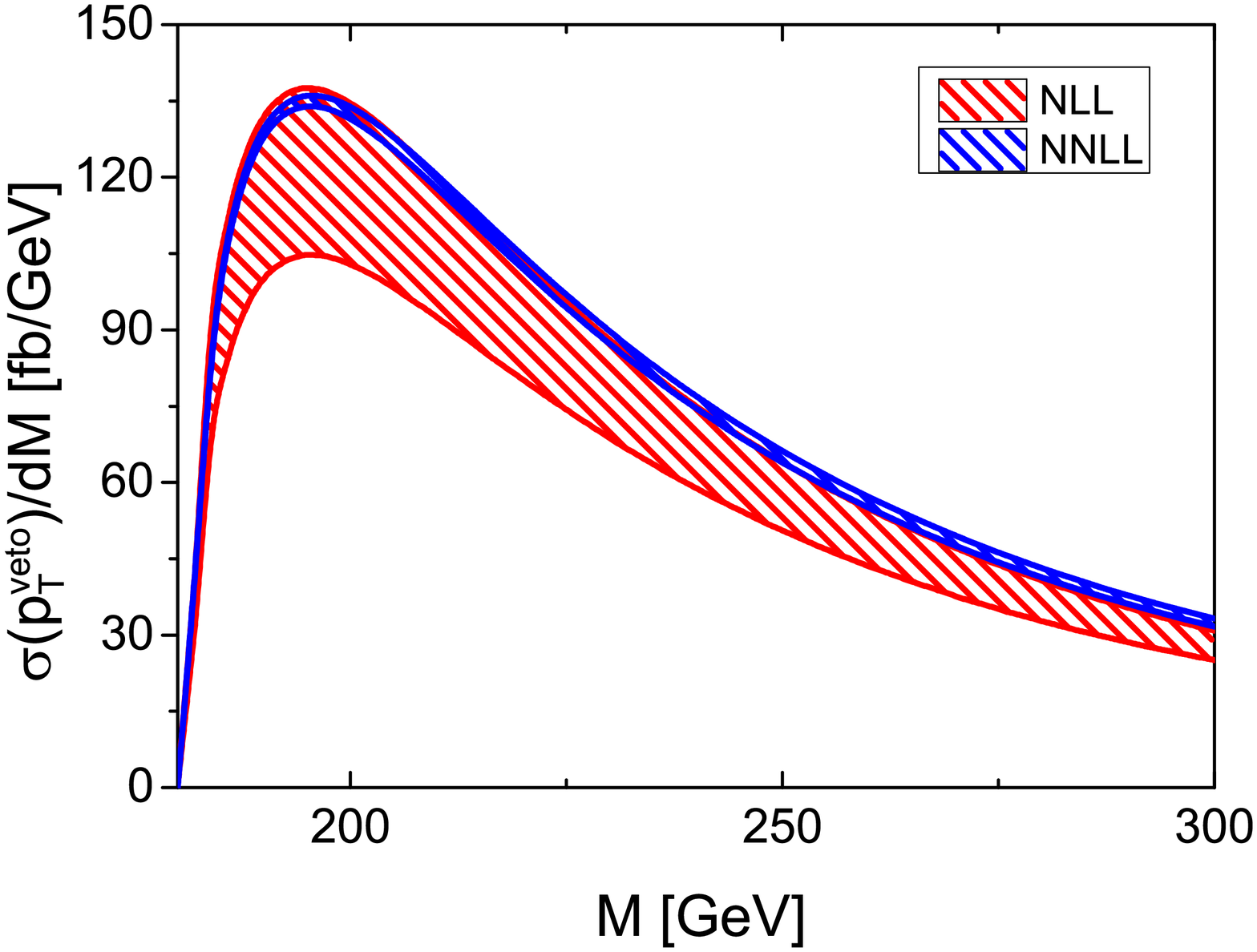}\\
\end{minipage}
\caption{The invariant mass distributions with scale uncertaintie at the LHC with $\fos$. The left figure is for $\wz$ productions and the right one is for $\zz$ productions.}\label{f_inm_s2}
\end{figure}

In Fig.~\ref{f_inm_s2}, we show the invariant mass distributions with factorization scale uncertainties for $\wz$ and $\zz$  production at the LHC, where we choose $\ptv =30~\gev$, $R=0.4$ at $\fos$, and $\pis$ are also included.
In all invariant mass region, the scale uncertainties are reduced obviously from NLL to NNLL level.

\subsection{RG improved predictions}
\begin{figure}[t!]
\begin{minipage}{0.49\linewidth}
\centering
  \includegraphics[width=1.0\linewidth]{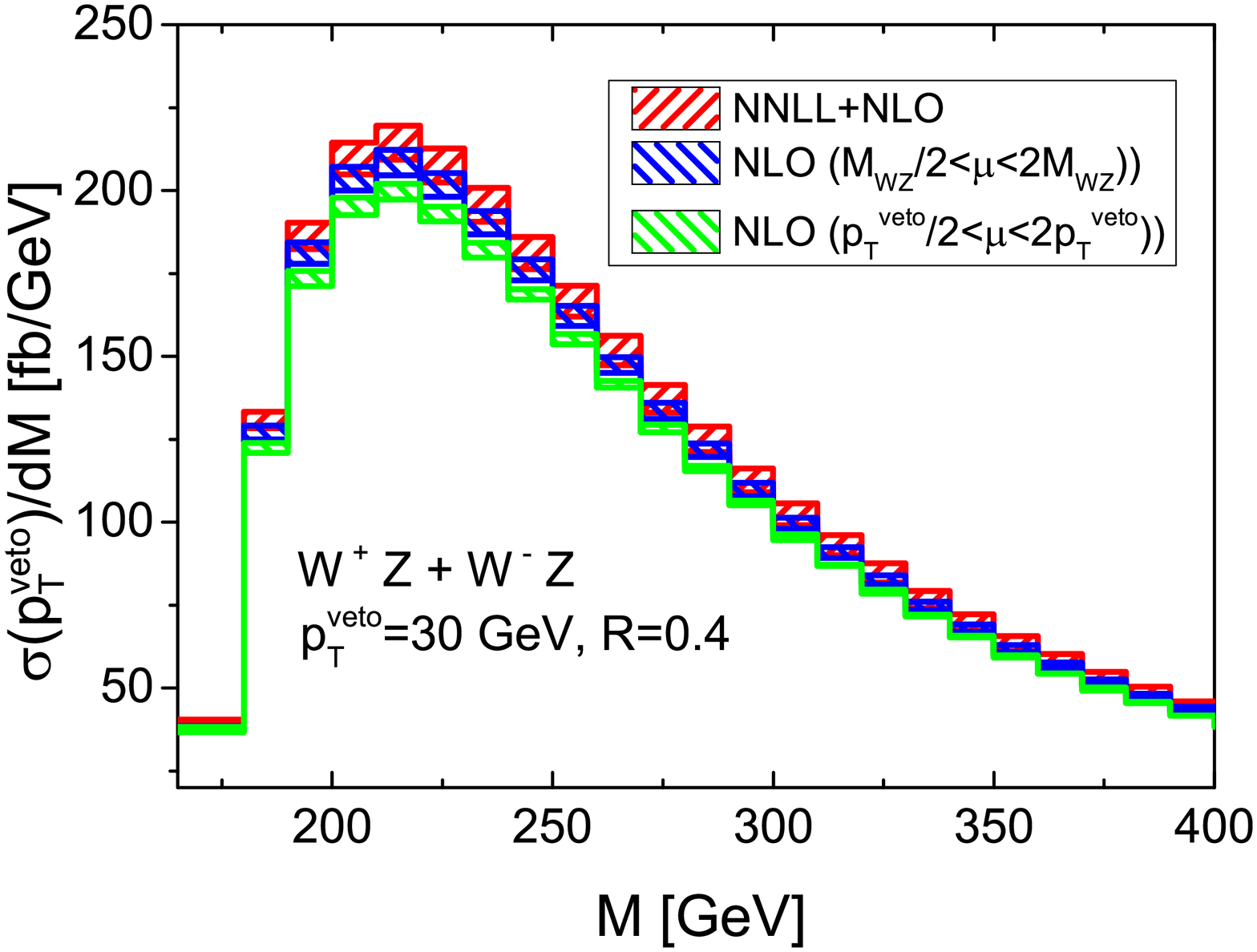}\\
\end{minipage}
\hfill
\begin{minipage}{0.49\linewidth}
\centering
 \includegraphics[width=1.0\linewidth]{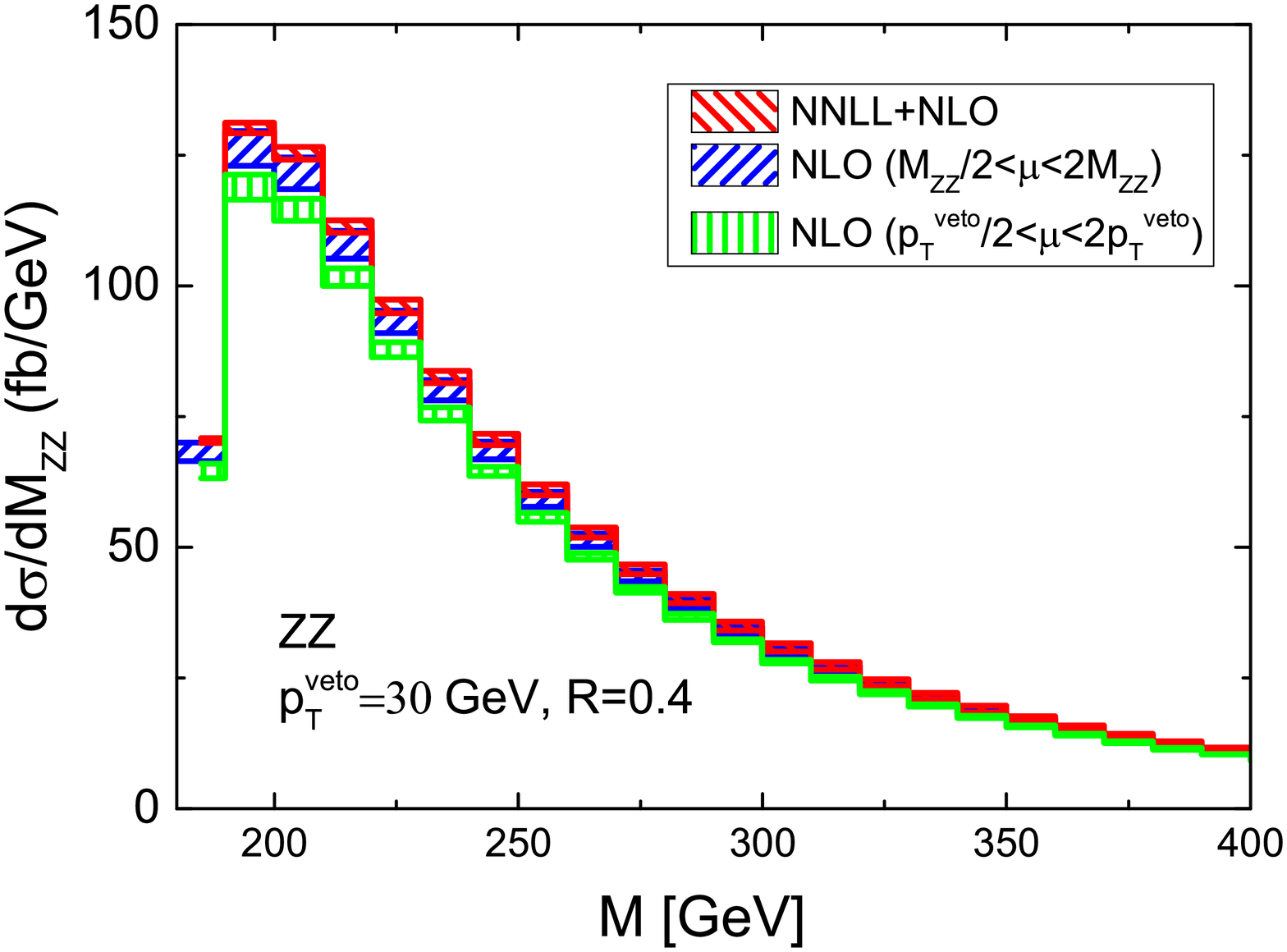}\\
\end{minipage}
\caption{The $\nlonnll$ invariant mass distributions with scale uncertaintie for $\wz$ and $\zz$ production at the LHC with $\ptv = 30~\gev$ and $R = 0.4$ with $\fos$. The left figure is for $\wz$ productions and the right one is for $\zz$ productions.}\label{f_inm_s3}
\end{figure}

In this section, we perform the complete $\nlonnll$ resummation results by Eq.~\ref{s2_eq_fullcs}.
In Fig.~\ref{f_inm_s3}, we show the invariant mass distributions for $\wz$ and $\zz$ production with $\pis$ at the LHC when $\fos$. The $gg$ channel for $\zz$ production are also included. The NLO results are presented in two benchmark scale schemes, $\mu_f \sim \mvv$ and $\mu_f \sim \ptv$, respectively. The uncertainties, in the case of $\mu_f \sim \ptv$, mainly concentrate in the large invariant mass region, and the differential cross sections are smaller than the resummation results. And for $\mu_f \sim \mvv$, the NLO differential cross sections are similar to the resummation results, but suffering a large scale uncertainties in all invariant mass region.

\begin{figure}[t]
\begin{minipage}{0.49\linewidth}
\centering
  \includegraphics[width=1.0\linewidth]{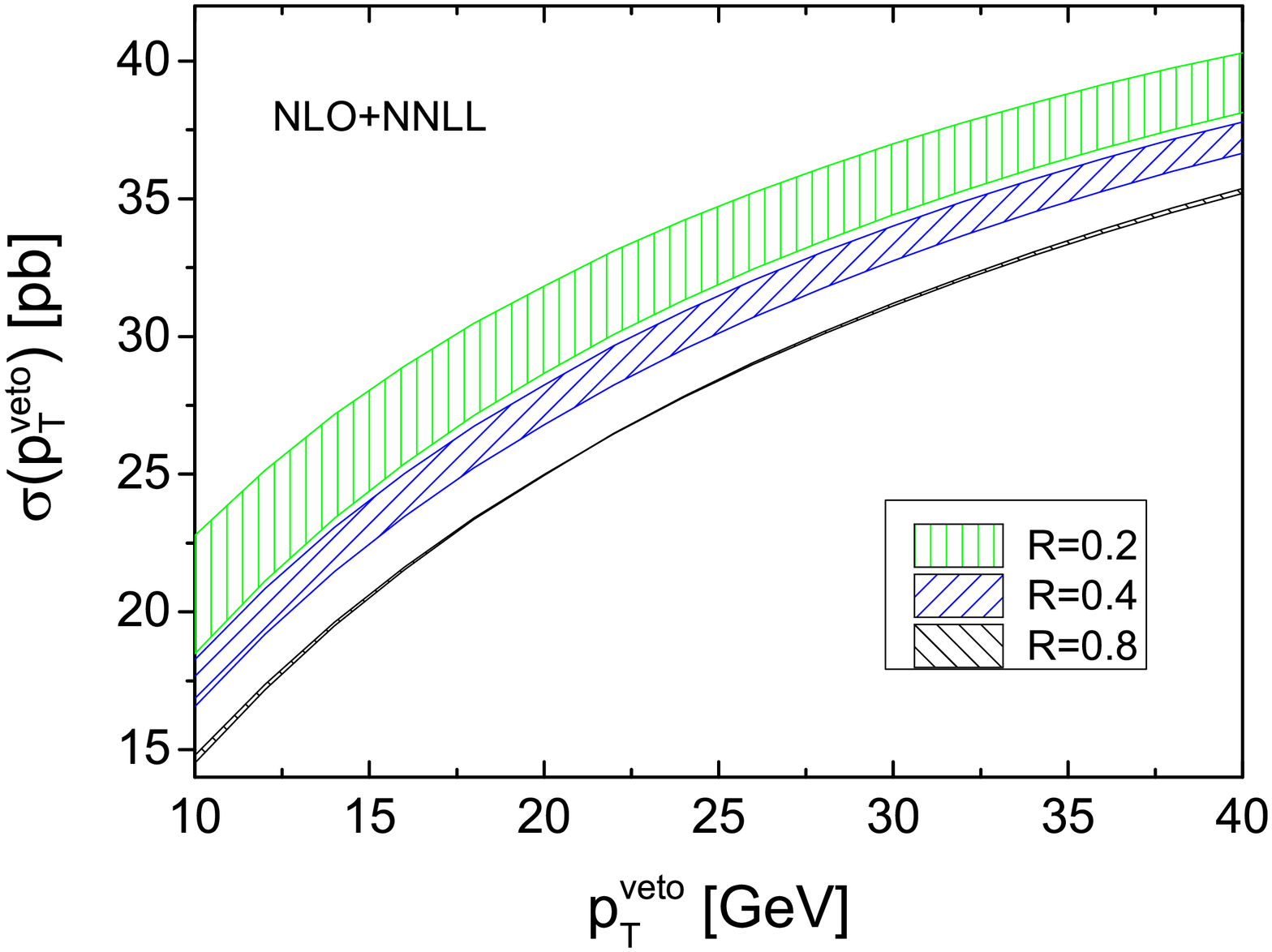}\\
\end{minipage}
\hfill
\begin{minipage}{0.49\linewidth}
\centering
 \includegraphics[width=1.0\linewidth]{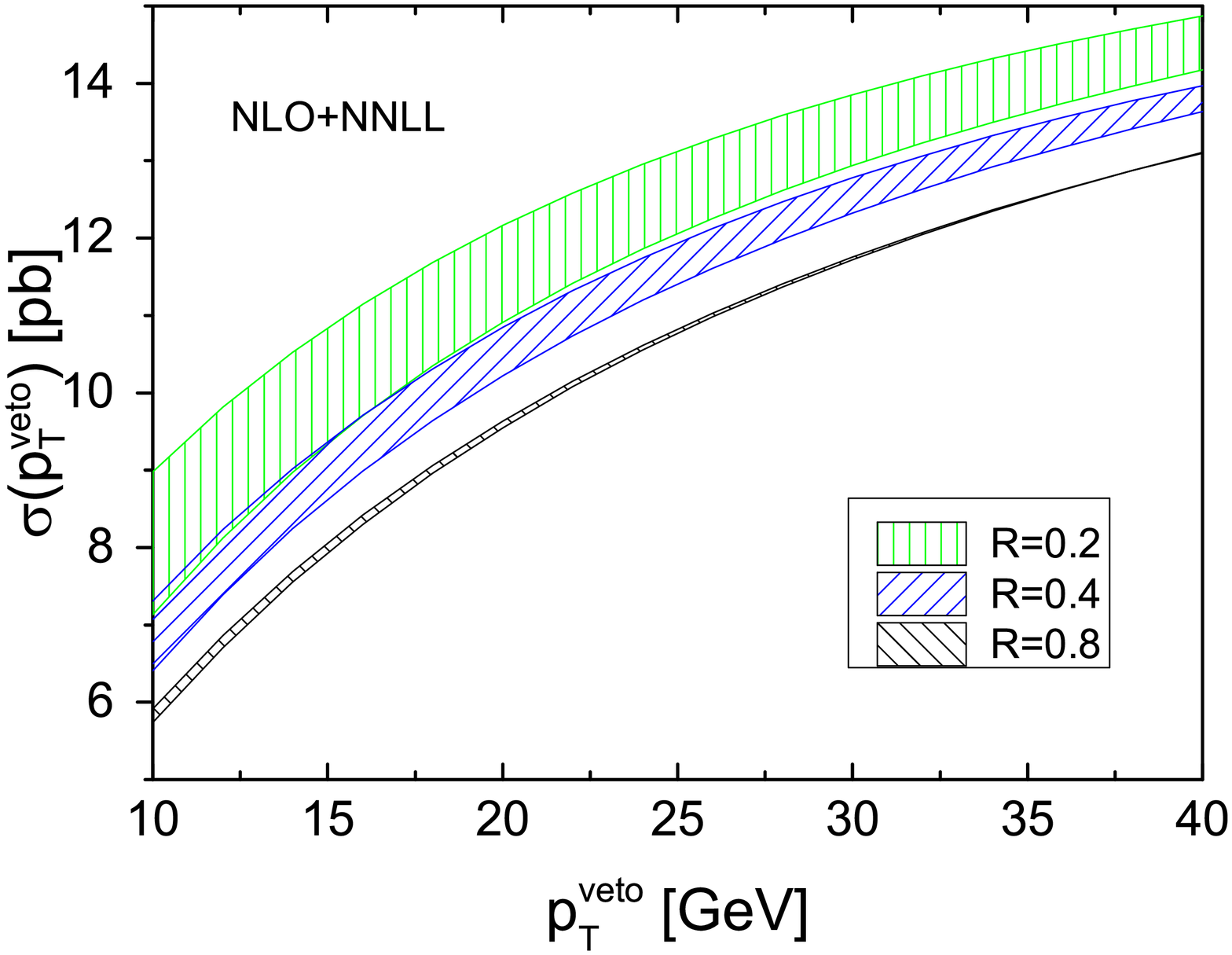}\\
\end{minipage}
\caption{The $\nlonnll$ total cross sections for different R  with $\fos$. The left figure is $\wz$ production and the right one is $\zz$ production.}\label{f_R_s3}
\end{figure}

In Fig.~\ref{f_R_s3}, we show the jet-veto cross sections for $R = 0.2,~0.4$ and $0.8$ at $\fos$. As stated before, the resummation predictions strongly depend on the jet radius. With the increasing of R, the $\nlonnll$ resummation results decrease, and the scale uncertainties are also reduced.
Fig.~\ref{f_R_s3} shows for $R=0.2, 0.4$, $0.8$ and $\ptv=30~\gev$, the factorization scale uncertainties are $7.2\%$, $3.8\%$ and $0.30\%$ for $\wz$ productions, and $6.8\%$, $3.6\%$ and $0.28\%$ for $\zz$ productions, respectively.

\begin{figure}[t]
\begin{minipage}{0.48\linewidth}
\centering
  \includegraphics[width=1.0\linewidth]{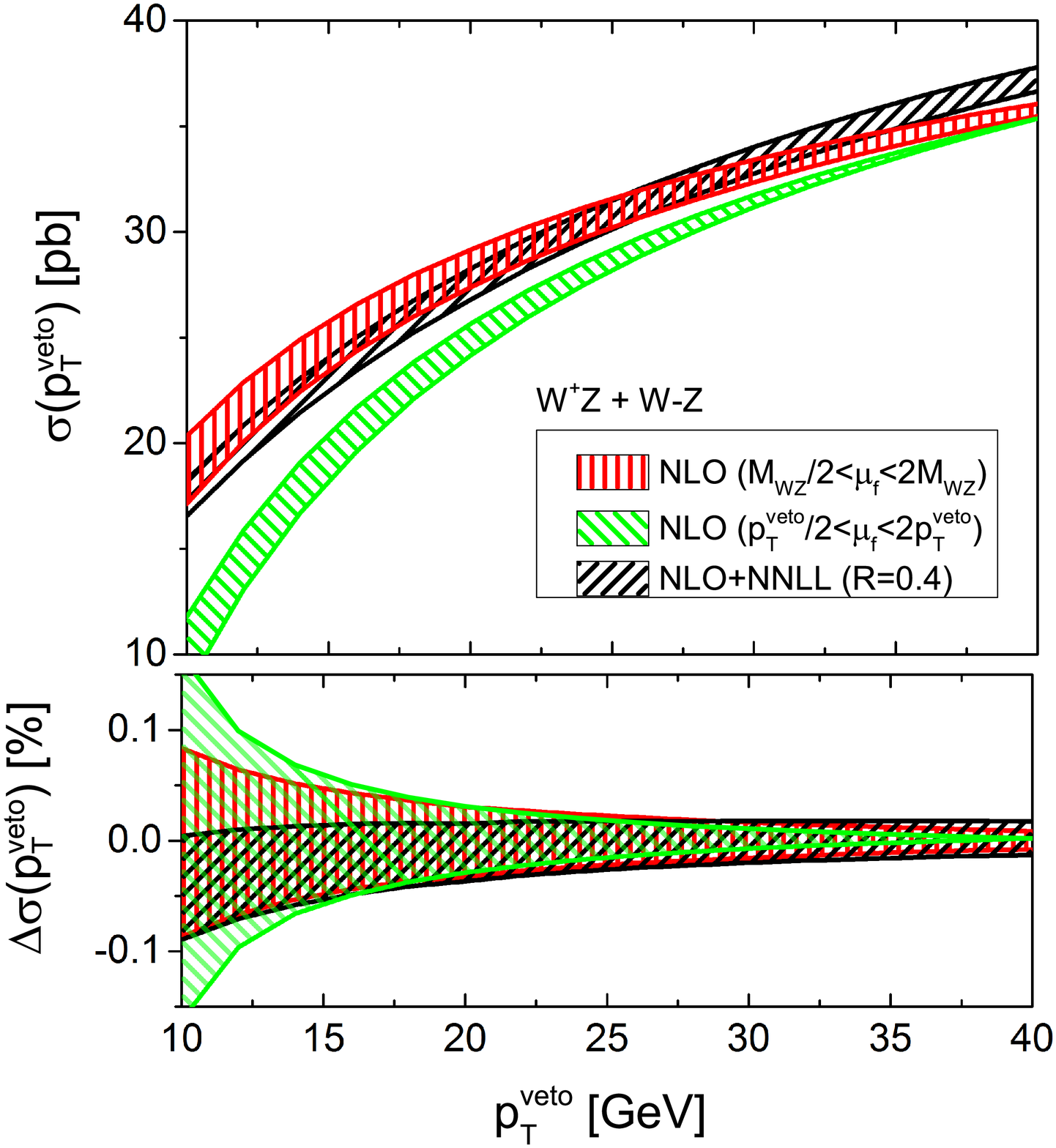}\\
\end{minipage}
\hfill
\begin{minipage}{0.48\linewidth}
\centering
 \includegraphics[width=1.0\linewidth]{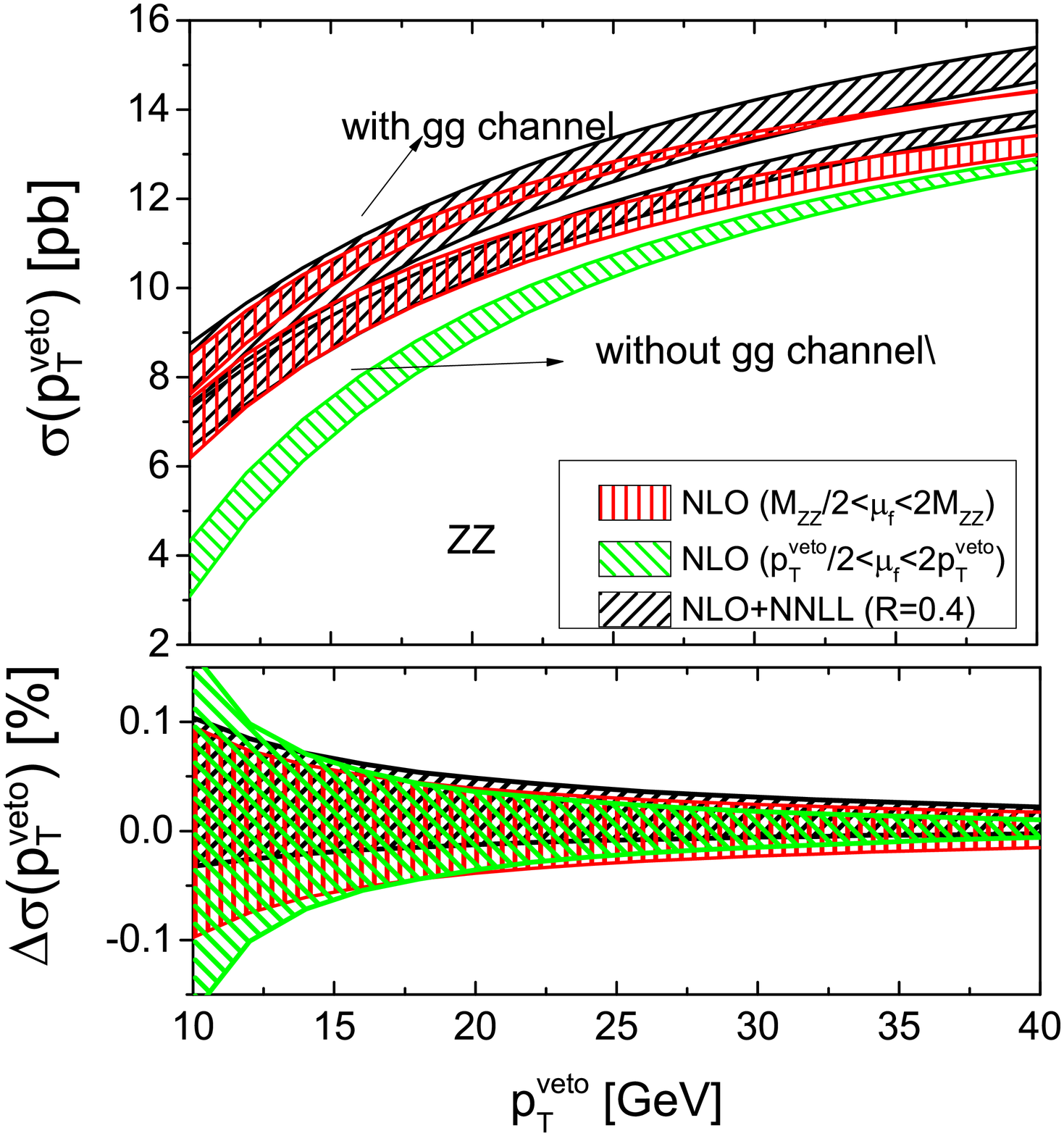}\\
\end{minipage}
\caption{Comparing the NLO and the $\nlonnll$ predictions at $\fos$. The left figure is for $\wz$ productions and the right one is for $\zz$ productions.}\label{f_nlo_s3}
\end{figure}

In Fig.~\ref{f_nlo_s3}, the comparison of the NLO and $\nlonnll$ jet-vetoed resummation is presented when $R=0.4$. The dependence of factorization scales in the NLO results are presented in two schemes, i.e. $\mu_f \sim \mvv$ and $\mu_f \sim \ptv$. From Fig.~\ref{f_nlo_s3}, in the small $\ptv$ region, the uncertainties of the NLO results are larger than the $\nlonnll$ predictions, especially for the choice of $\mu_f \sim \ptv$. It reflects that the fix-order results are unbelievable in the small $\ptv$ region, while the resummation make it reliable. When $\ptv \succeq30~\gev$, the jet-veto resummed cross section become larger than the NLO results for $\mu_f \sim \mvv$.

\subsection{Comparison with the experiment}\label{s34}
\begin{figure}[t]
\begin{minipage}{0.49\linewidth}
\centering
  \includegraphics[width=1.0\linewidth]{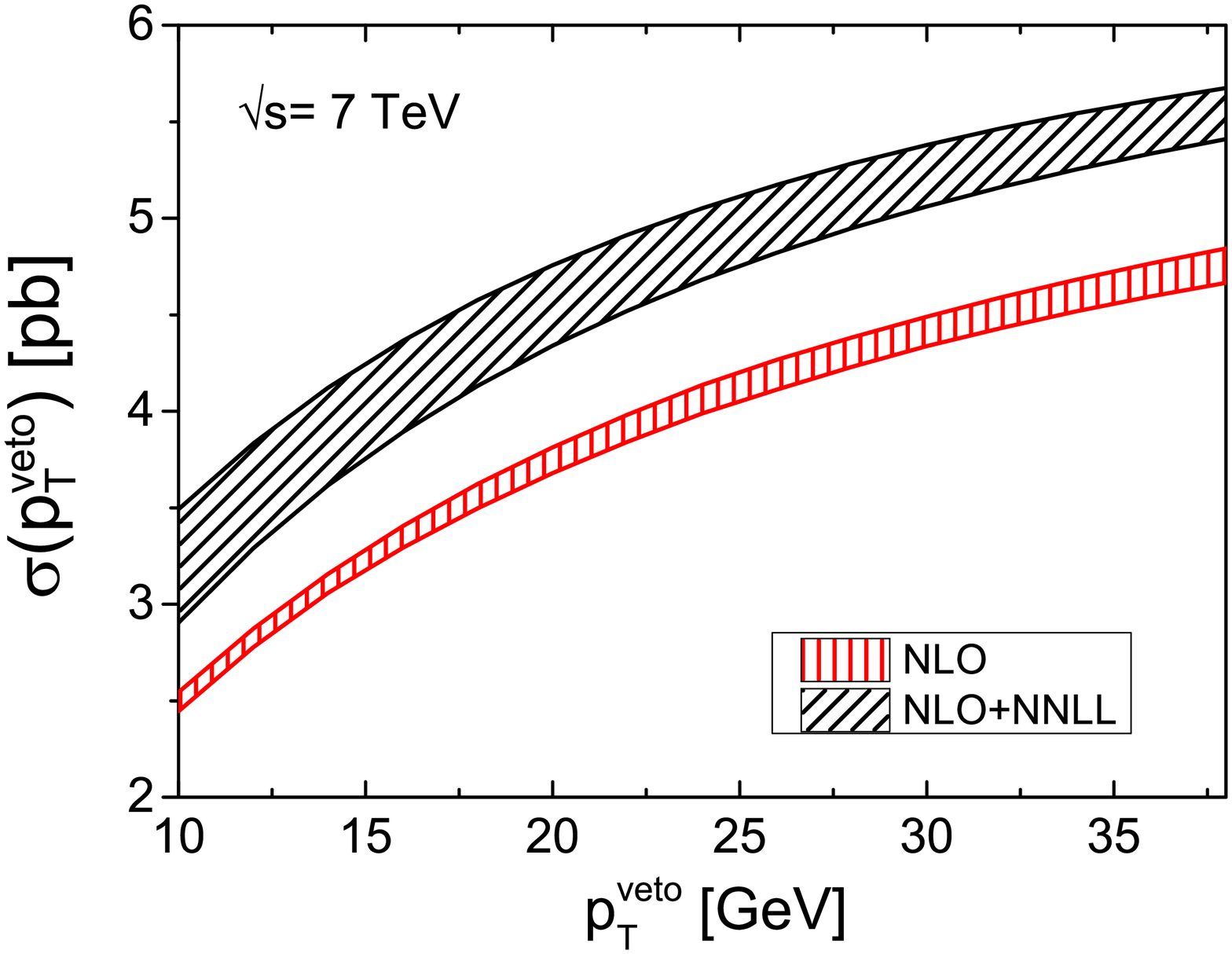}\\
\end{minipage}
\hfill
\begin{minipage}{0.49\linewidth}
\centering
  \includegraphics[width=1.0\linewidth]{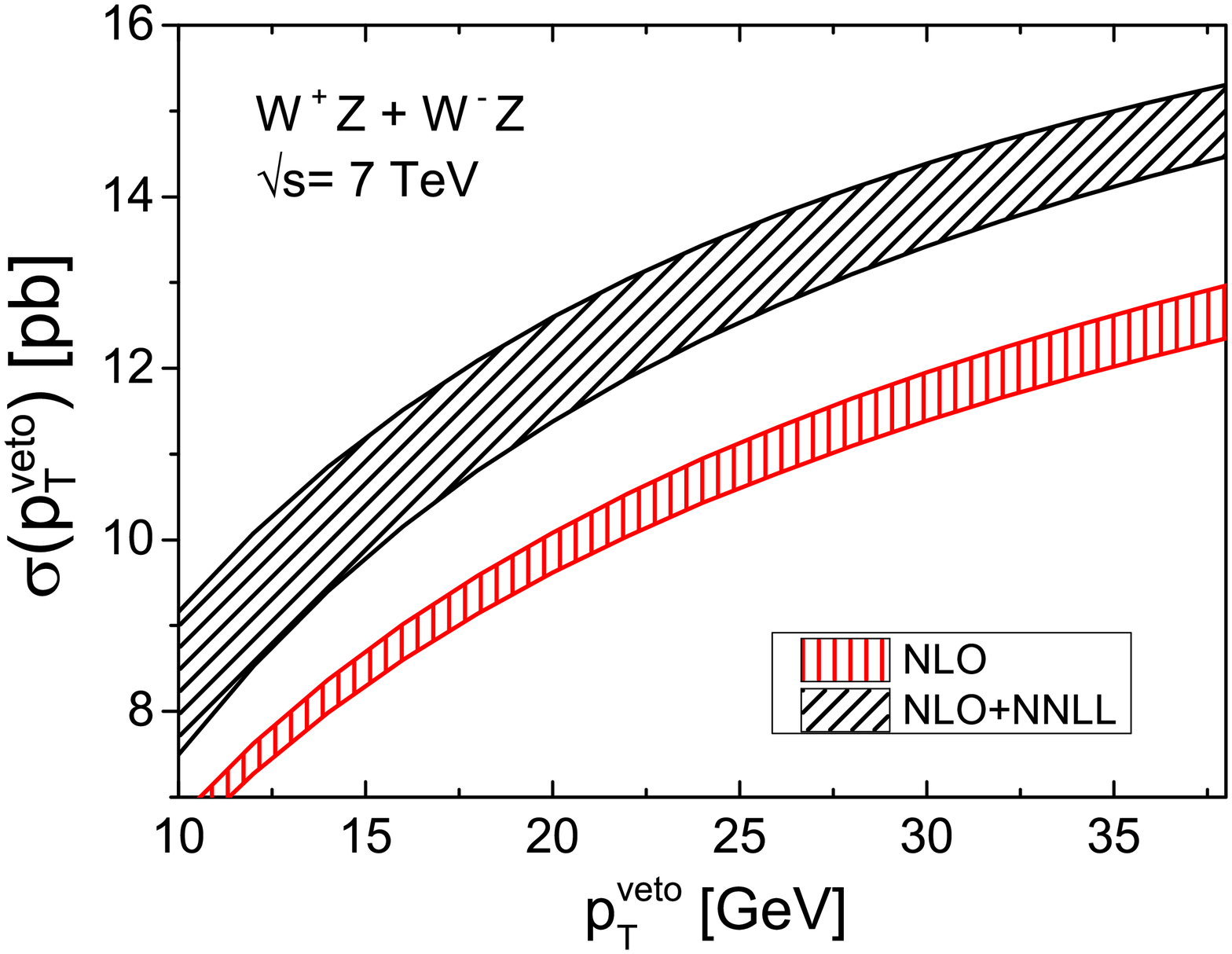}\\
\end{minipage}
\hfill
\caption{Comparing the POWHEG+PYTHIA results and the $\nlonnll$ predictions for $\wz$ and $\zz$ production at $\ses$, where $R=0.4$.}\label{f_eff_7tev_s3}
\end{figure}

\begin{figure}[t]
\begin{minipage}{0.49\linewidth}
\centering
  \includegraphics[width=1.0\linewidth]{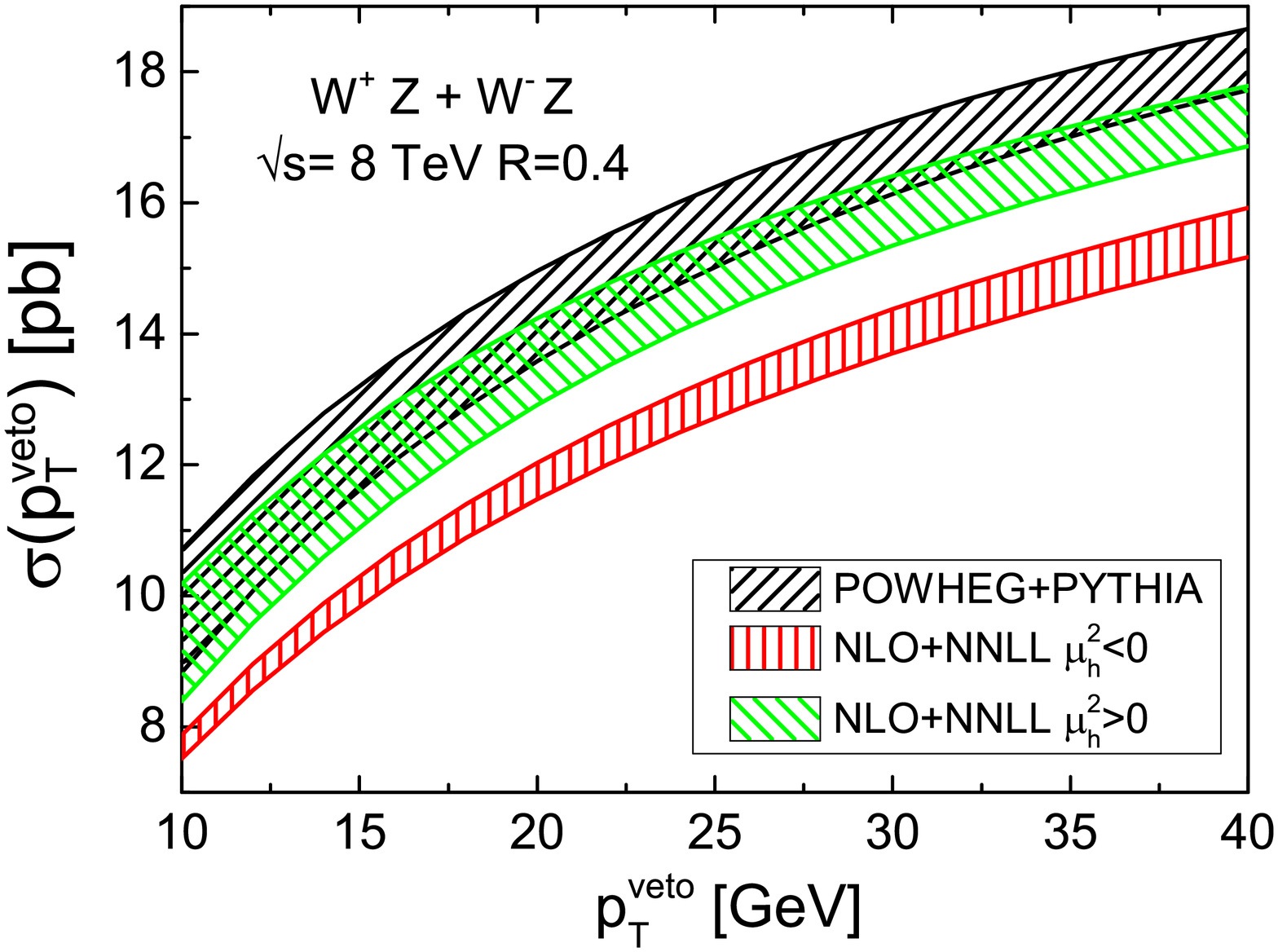}\\
\end{minipage}
\hfill
\begin{minipage}{0.49\linewidth}
\centering
  \includegraphics[width=1.0\linewidth]{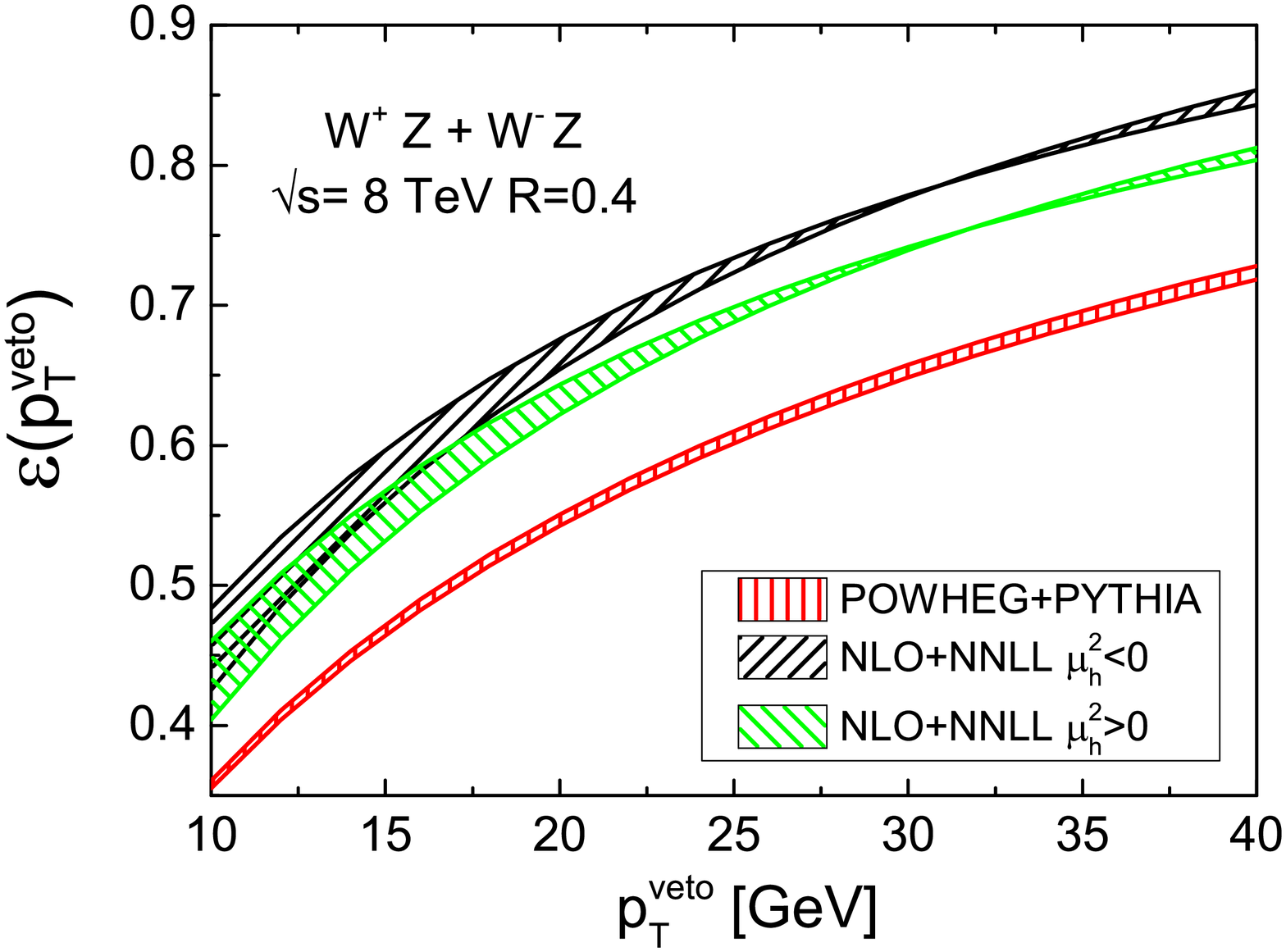}\\
\end{minipage}
\hfill
\begin{minipage}{0.49\linewidth}
\centering
  \includegraphics[width=1.0\linewidth]{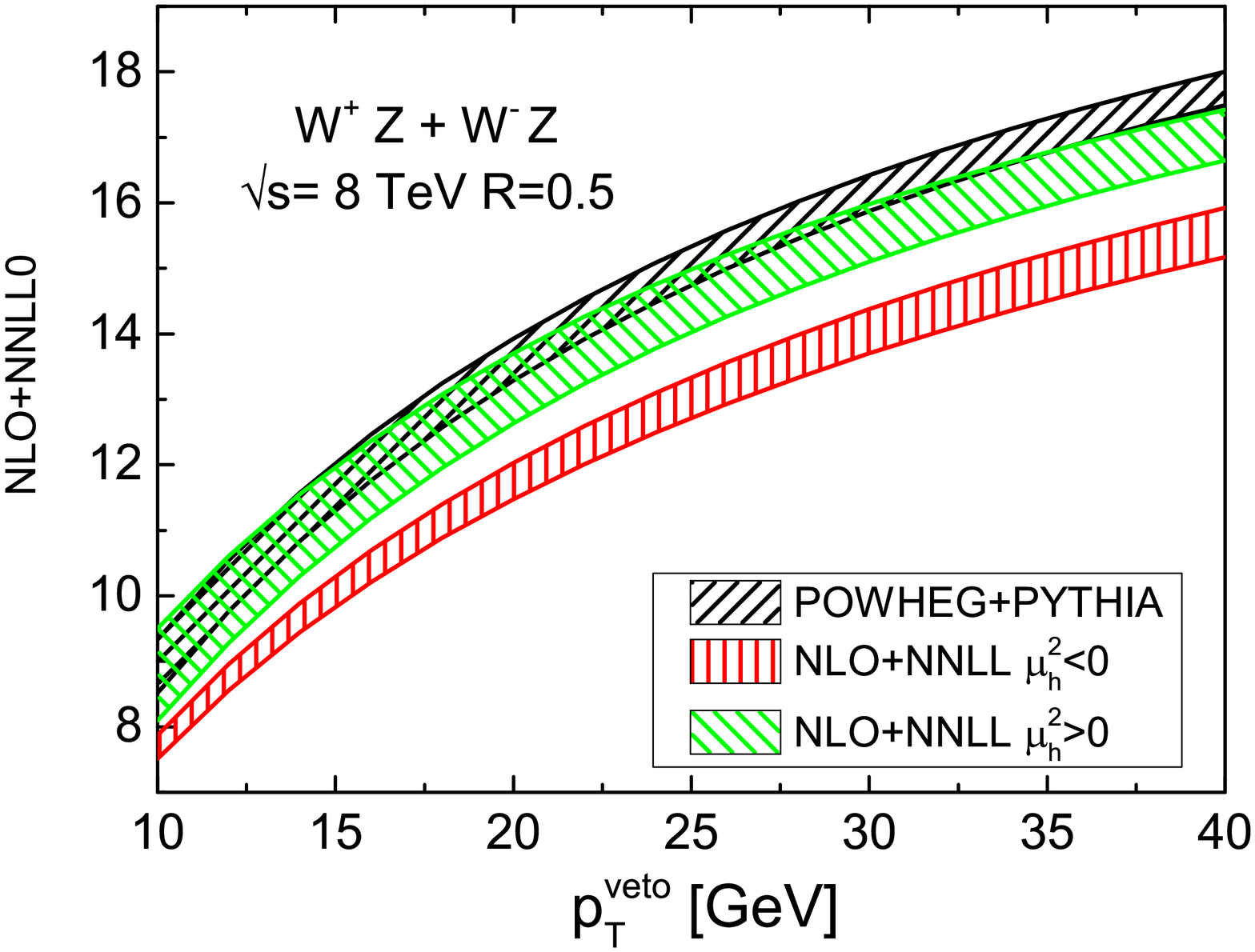}\\
\end{minipage}
\hfill
\begin{minipage}{0.49\linewidth}
\centering
  \includegraphics[width=1.0\linewidth]{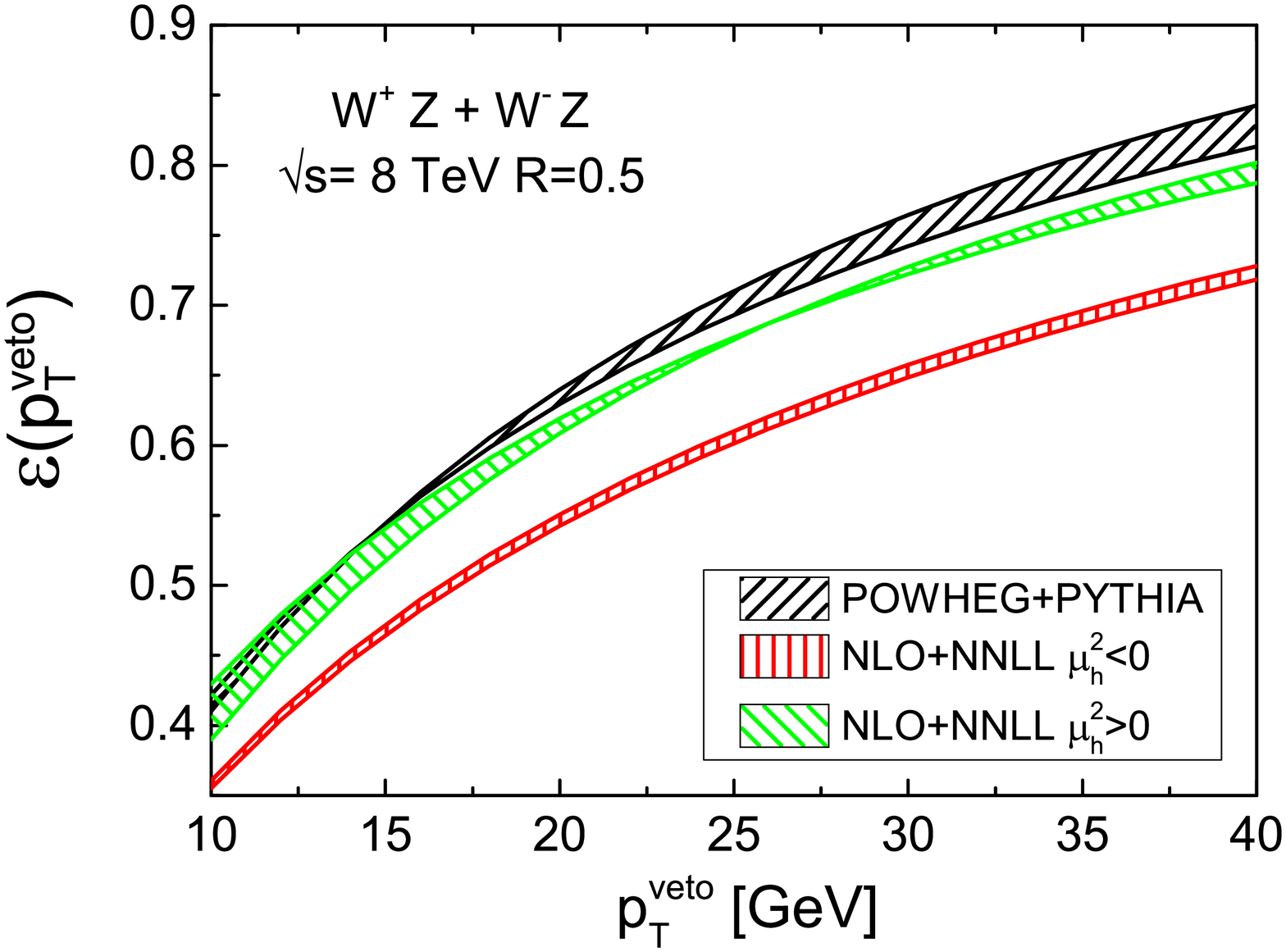}\\
\end{minipage}
\hfill
\caption{Comparing the POWHEG+PYTHIA results and the $\nlonnll$ predictions, as well as their efficiencies, for $\wz$ production with $R=0.4$ and $R=0.5$ at $\eis$, respectively.}\label{f_eff_wz_s3}
\end{figure}

\begin{figure}[t]
\begin{minipage}{0.49\linewidth}
\centering
  \includegraphics[width=1.0\linewidth]{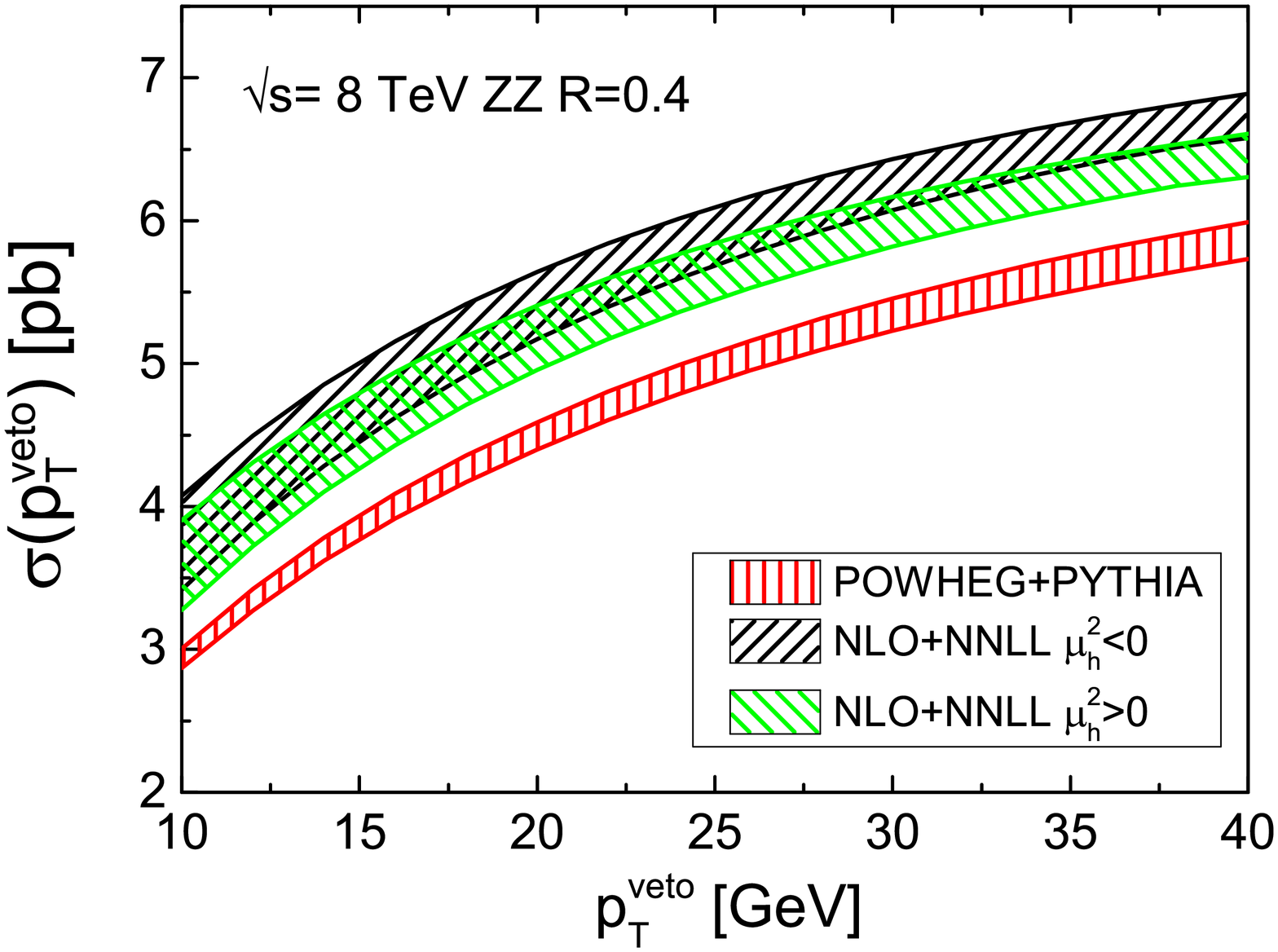}\\
\end{minipage}
\hfill
\begin{minipage}{0.49\linewidth}
\centering
  \includegraphics[width=1.0\linewidth]{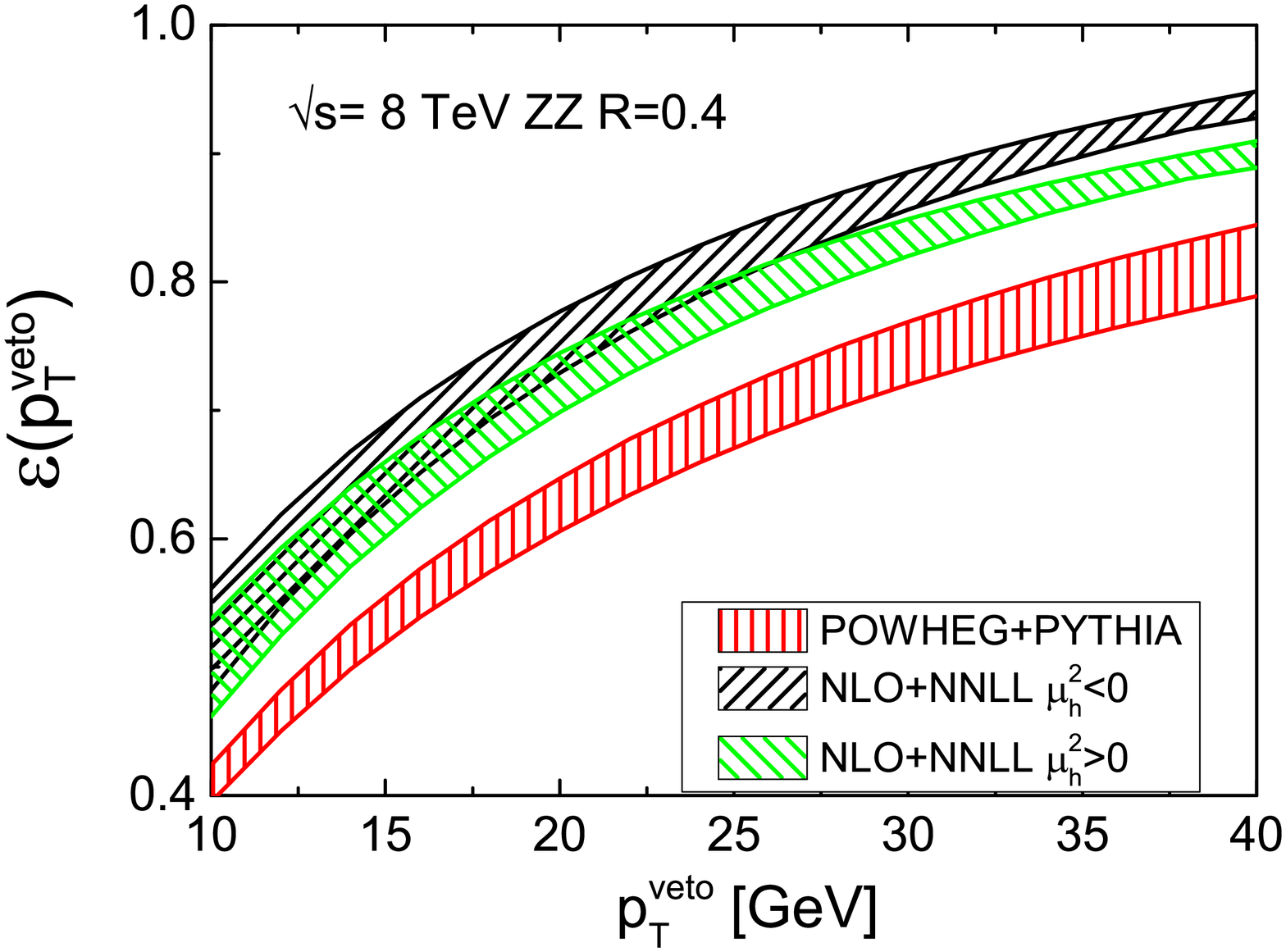}\\
\end{minipage}
\hfill
\begin{minipage}{0.49\linewidth}
\centering
  \includegraphics[width=1.0\linewidth]{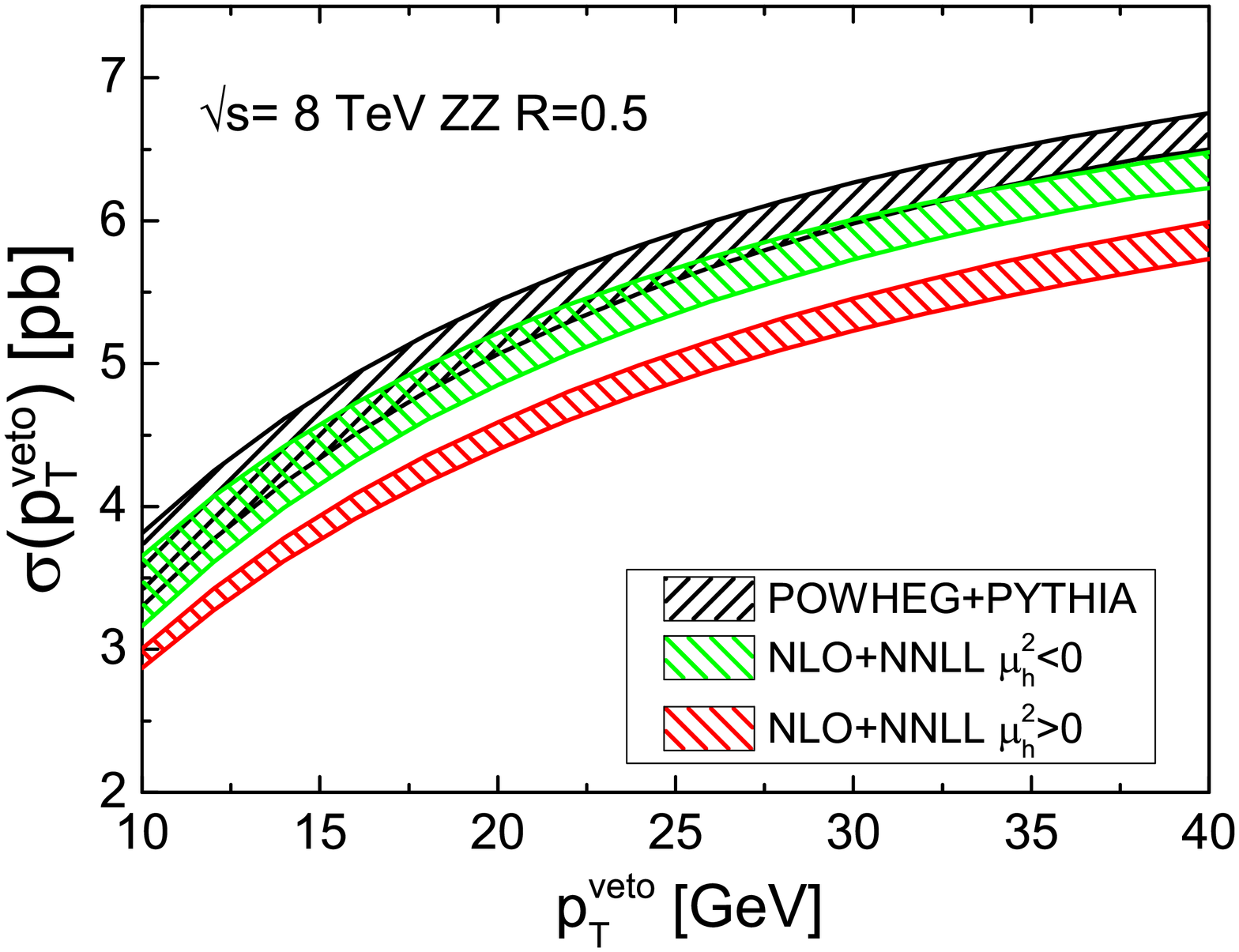}\\
\end{minipage}
\hfill
\begin{minipage}{0.49\linewidth}
\centering
  \includegraphics[width=1.0\linewidth]{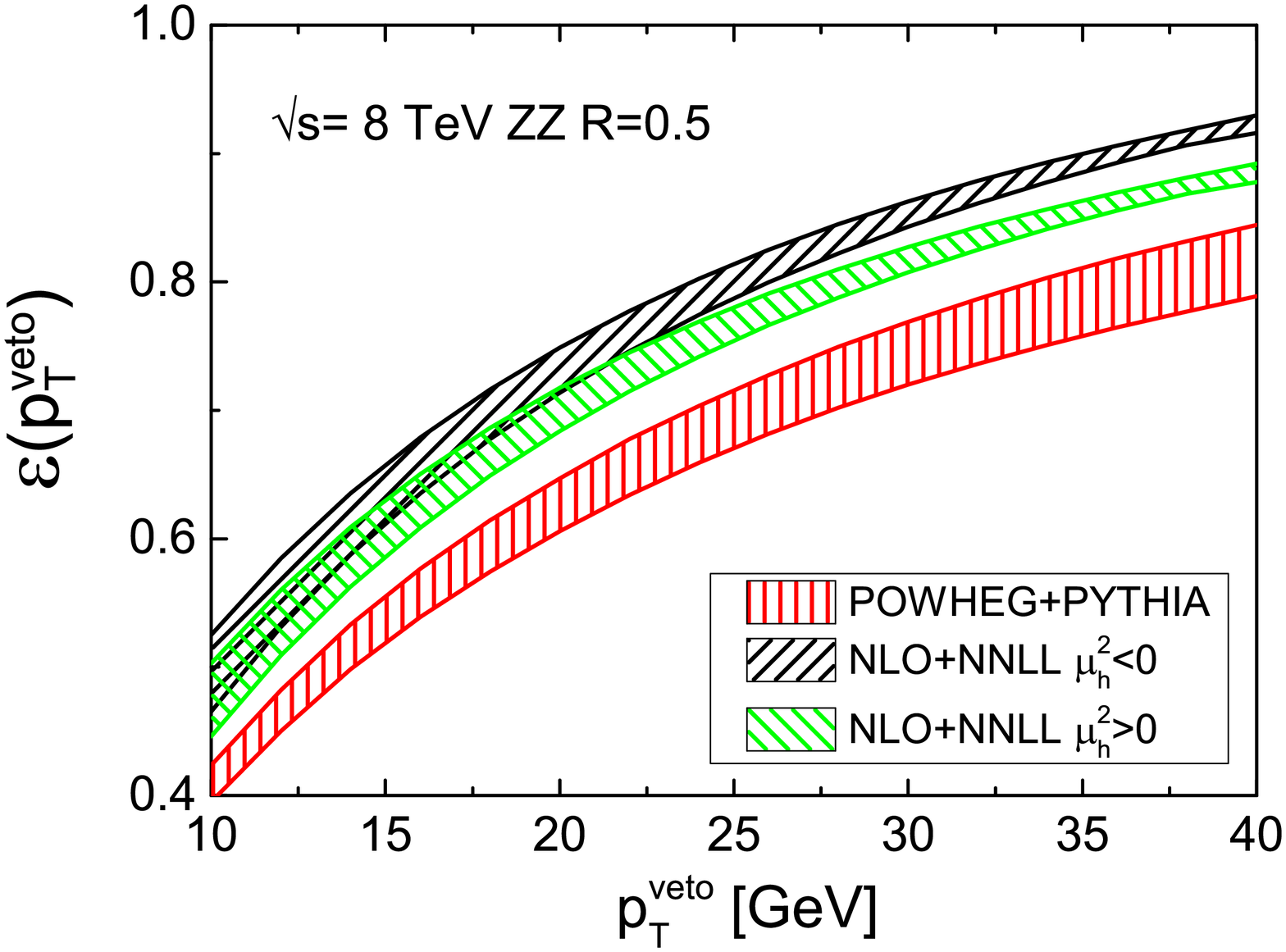}\\
\end{minipage}
\hfill
\caption{Comparing the POWHEG+PYTHIA results and the $\nlonnll$ predictions, as well as their efficiencies, for $\zz$ production with $R=0.4$ and $R=0.5$ at $\eis$, respectively }\label{f_eff_zz_s3}
\end{figure}

We also compare our resummed results and jet-veto efficiencies with those calculated by the POWHEG+PYTHIA~\cite{Melia:2011tj} in In  Fig.~\ref{f_eff_7tev_s3}, Fig.~\ref{f_eff_wz_s3} and Fig.~\ref{f_eff_zz_s3}. The $\nlonnll$ bands are obtained by varying the hard and factorization scales by factors of 2 about their default values, and the NLO bands reflect the renormalization and factorization scale uncertainties for the  POWHEG+PYTHIA results, respectively.

In  Fig.~\ref{f_eff_7tev_s3}, the $\nlonnll$ results  include $\pi^2$ resummations with $R=0.4$ at $\ses$. When $\ptv=30~\gev$, $\nlonnll$ calculations increase the NLO results by about $19\%$ ($18\%$), and the jet-veto efficiencies are increased by about $13\%$ ($13\%$) for $\wz$ ($\zz$) productions.

In  Fig.~\ref{f_eff_wz_s3} and Fig.~\ref{f_eff_zz_s3}, we present comparisons  between POWHEG+PYTHIA results and the $\nlonnll$ predictions when $R=0.4$ and $R=0.5$ for $\wz$ and $\zz$ productions, respectively. We also include results without $\pi^2$ enhancement effects in each figures.  In general, our resummed cross sections and jet-veto efficiencies, no matter including $\pi^2$ effects (${\muhs}<0$) or not (${\muhs}>0$), are larger than POWHEG+PYTHIA predictions.
When $R=0.4$ and $R=0.5$, the jet-veto cross section with (without) $\pi^2$ enhancement effects are increased by about $19\%$ ($13\%$) and $15\%$  ($11\%$) for $\ptv=30~\gev$ for $\wz$ production,  $18\%$ (13\%) and $15\%$ ($11\%$) for $\zz$ production, respectively.
As shown above, it also can be seen that the scale uncertainties for $\nlonnll$ are better than the POWHEG+PYTHIA results for $R=0.5$, but when $R=0.4$,  the scale uncertainties of resummed results are a little larger.
Comparing with the Fig~\ref{f_nlo_s3}, Fig~\ref{f_eff_7tev_s3} and Fig~\ref{f_eff_wz_s3}, which correspond different center-of-mass energy, we can find that the scale uncertainties of the resummed results reduce significantly with the increasing of the center-of-mass energy. When $\sqrt{S}=7,8$ and $14$ TeV, the scale uncertainties for $\wz$ productions are $6.9\%$, $6.4\%$, $3.8\%$ for $\wz$ production, and $6.1\%$, $5.6\%$, $3.6\%$ for $\zz$ production, respectively, for $R=0.4$, $\ptv=30\gev$, where $\pi^2$ enhancement effects are included.

\begin{figure}[t!]
  \includegraphics[width=0.7\linewidth]{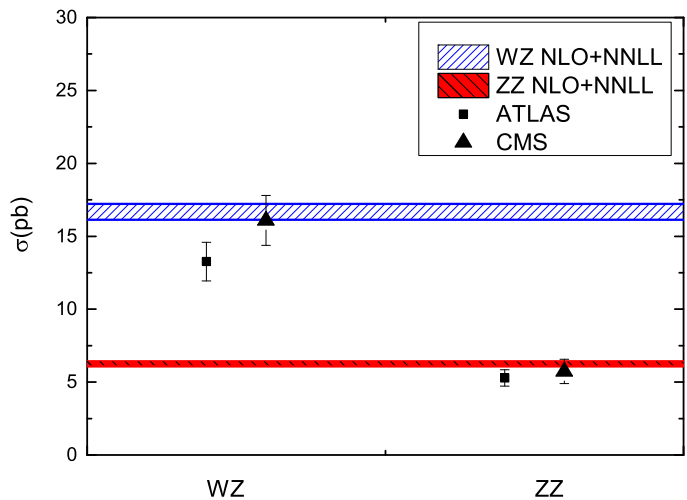}\\
  \caption{Comparison of total cross sections for $\wz$ and $\zz$ productions between experimental data and resummation prediction at the LHC with $\sqrt{S}=8$ TeV.}\label{f_exp}
\end{figure}

In Table~\ref{t_exp_s3}, we show the comparison of experiment data measured by the ATLAS and CMS Collaboration with our jet-veto resummed predictions at $\eis$. To obtain the experimental jet-vetoed cross sections, we use experimental total cross section multiply the jet-veto efficiency factors calculated from the POWHEG+PYTHIA simulation~\cite{Jaiswal:2014yba}, which are also presented in the Fig.~\ref{f_exp}. From the Fig.~\ref{f_exp}, we can see that our predictions agree with the CMS experiments in 2$\sigma$ for both $\wz$ and $\zz$ productions, while there is discrepancy between the experimental results and NLO predictions for $\wz$ production. But our results are higher than the ATLAS jet-veto results. Actually, the ATLAS and CMS data is different at $eis$ so far, but we only compare their data against our theoretical results with the same parameters (i.e. $\ptv$ and R) here; another possible reason is both ATLAS and CMS collaborations use different tunes for parton shower generators, and we only used the default tunes.
 \begin{table}[t]
\begin{center}
\begin{tabular}{|c|c|c|c|}
    \hline
  ~~~~ & $ATLAS^*$ & $CMS^*$ & Theory   \\
    \hline
     \hline
  $\sigma^{veto}_{\wz}[pb]$              & $13.26\pm 1.33$      & $16.09\pm 1.70$     & $16.67\pm0.55$    \\
    $\sigma^{veto}_{\zz}[pb]$              & $5.28\pm0.56$      & $5.73\pm0.83$     & $6.25\pm0.18$\\
        \hline
\end{tabular}
\end{center}
  \caption{Comparison of experiment data from the ATLAS and CMS Collaboration
 with our jet-veto resummed cross sections for $\wz$  and $\zz$ production when $\eis$. The signs of $ATLAS^*$  and $CMS^*$ assign the cross section extrapolating by using reported experimental total cross section multiply the jet-veto efficiency of POWHEG+PYTHIA simulation.}\label{t_exp_s3}
\end{table}

\section{Conclusion}\label{s4}
We have calculated the jet-vetoed resummation for $\wz$  and  $\zz$ pair productions at the $\nlonnll$ accuracy with SCET at the LHC. We present the invariant mass distributions and the total cross sections, including $\pi^2$ enhancement effects. Our results show that the jet-veto resummation can increase the jet-veto cross section and decrease the scale uncertainties, especially in the large center-of-mass energy.
The resummation results, for $\ptv>30~\gev$ and $R=0.4$,can increase POWHEG+PYTHIA predictions by about $19\%$ for $\wz$ production and $18\%$ for $\zz$ production, respectively. In $\wz$ channel our resummed results agree with CMS experiment data  within 2$\sigma$ C.L. at $\eis$, which can explain the 2$\sigma$ discrepancy found between the  CMS experimental results and theoretical predictions based on NLO calculation with parton showers. .

\section{Acknowledgments}
This work was supported in part by the National Natural Science Foundation of China under Grants No. 11375013 and No. 11135003.

\bibliography{wz}
\end{document}